\documentclass[aps,prl,twocolumn,groupedaddress,floatfix,nofootinbib]{revtex4-2}
\usepackage{amsmath,amssymb,graphicx,bm}
\usepackage[colorlinks=true,linkcolor=blue,citecolor=blue,urlcolor=blue]{hyperref}
\graphicspath{{figures/}}
\newcommand{\ellK}{\mathcal{I}}

\begin{document}

\title{Kinetic competition controls loss-induced exceptional points\\
of Hubbard pairs}

\author{Ho-Kin Tang\textsuperscript{1,3,}}
\email[Corresponding author: ]{denghaojian@hit.edu.cn}
\author{Xiaosen Yang\textsuperscript{2,}}
\email[Corresponding author: ]{yangxs@ujs.edu.cn}
\affiliation{\textsuperscript{1}School of Science, Harbin Institute of Technology, Shenzhen 518055, China\\
\textsuperscript{2}Department of Physics, Jiangsu University, Zhenjiang 212013, China\\
\textsuperscript{3}Shenzhen Key Laboratory of Advanced Functional Carbon Materials
Research and Comprehensive Application, Shenzhen 518055, China}
\date{\today}

\begin{abstract}
Two-body loss can bind two fermions in a Hubbard lattice, yet the binding
threshold in the infinite lattice need not be an exceptional point (EP). At
fixed total momentum $\bm K$, the hops of the two particles interfere into a
relative-motion hopping along each lattice direction, which sets the kinetic
energy of the relative motion. On a hypercubic lattice, the threshold
$g_c=1/[\pi\rho_2(0)]$ is an EP precisely when the maximum kinetic energy along
one direction exceeds the sum of the transverse maxima. This dominance keeps
every transverse-momentum scattering channel propagating near the band center,
making the local pair Green function purely imaginary there. Two real-energy
scattering zeros associated with coherent perfect absorption then coalesce at
the threshold into a second-order EP of the analytically continued pair
problem, where loss restores the parity-time symmetry of the pair's
scattering solutions. Without a dominant direction, binding occurs without
coalescence and without breaking of that symmetry.
Total momentum can therefore select EP sectors even on isotropic lattices.
More broadly, the kinetic structure of the two-particle continuum determines
whether loss-induced binding and exceptional degeneracy coincide.
\end{abstract}

\maketitle

\emph{Introduction.}---Exceptional points (EPs), where eigenvalues and
eigenvectors coalesce, are singularities of non-Hermitian
spectra~\cite{Kato1966,Berry2004,Heiss2012,ElGanainy2018,MiriAlu2019,Bergholtz2021,Ashida2020}
and have been realized in optics and quantum
circuits~\cite{Ruter2010,Peng2014,Naghiloo2019}. Two-body loss in a Hubbard lattice acts
as an imaginary on-site interaction in the no-jump dynamics~\cite{Daley2014}
and is accessible in cold-atom
gases~\cite{Syassen2008,Zhu2014,Tomita2017,Sponselee2019,Takasu2020,Honda2023}.
Such loss can bind two fermions into a decaying pair, but the emergence of a
bound state from the continuum does not itself imply an EP.
In interacting systems such non-Hermitian terms drive transitions without a
Hermitian counterpart~\cite{Nakagawa2018,Yamamoto2019,Hamazaki2019,Sun2022,Lu2024}
and produce EPs of few- and many-body
spectra~\cite{Luitz2019,Pan2019,Schafer2022,Ding2022,Wang2023HH}. Here the
non-Hermiticity enters through the interaction, not through the single-particle
bands of non-Bloch band
theory~\cite{Yao2018,Song2019,Xiao2020,Fu2022,Wang2024}. We determine when
the binding threshold also marks the coalescence of analytically continued
scattering states.

Coherent perfect absorption (CPA) occurs when a scattering-matrix eigenvalue
vanishes at real energy~\cite{Chong2010,Mostafazadeh2009}; we call such
energies CPA zeros. A perfectly absorbing EP arises
when two purely incoming solutions coalesce, giving a quartic minimum of
scattered intensity as a function of energy~\cite{Sweeney2019,Wang2021}.
For particles, complete pair annihilation has been found with a
one-dimensional imaginary contact interaction~\cite{LiZhangSong2014}, and
related lattice dynamics has been studied with imaginary
interactions~\cite{ZhangJinSong2017} and imaginary
hopping~\cite{ZhangSong2021}. Binding thresholds for a single imaginary
impurity are known on isotropic lattices~\cite{Kokkinakis2026}. These results
do not determine whether the pair-binding threshold is exceptional at fixed
total momentum in the infinite lattice.

\begin{figure}[!b]
\includegraphics[width=\columnwidth]{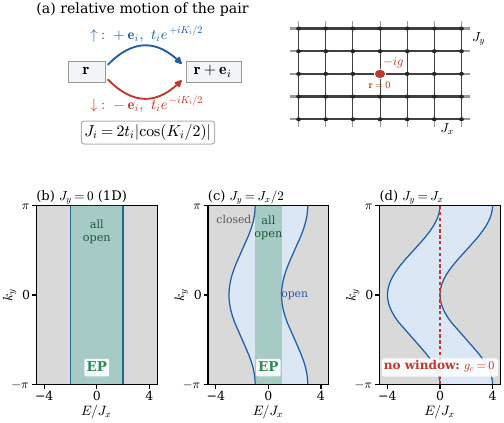}
\caption{Loss-induced binding as scattering of the relative motion. (a) At total
momentum $\bm K$ the relative coordinate $\mathbf r=\mathbf r_\uparrow-\mathbf r_\downarrow$
advances by $\mathbf e_i$ when the up fermion hops forward or the down fermion
backward; the amplitudes interfere to $2t_i\cos(K_i/2)$, whose magnitude is the
relative hopping $J_i$, and the loss is one
absorbing site, $-ig$, at $\mathbf r=0$. (b)--(d) At fixed $k_y$ the motion
along $x$ is a channel of half-width $2J_x$ shifted by $-2J_y\cos k_y$, open
(light blue) or closed (grey) at energy $E$. An EP needs a window in which every
channel is open (green), $|E|<2(J_x-J_y)$: (b) $J_y=0$, the whole band; (c)
$J_y=J_x/2$; (d) $J_y=J_x$, the window closes and there is no EP.}
\label{fig:channels}
\end{figure}
The distinction is controlled by the kinetic energy of relative motion. At
fixed total momentum, the two particles' hops combine differently along each
lattice direction~\cite{Mattis1986,Valiente2008,Winkler2006}. When the maximum kinetic-energy contribution along one
direction exceeds the sum of the transverse maxima, every transversely
shifted band includes energies near the center of the pair continuum~\cite{Economou2006}. Two
real-energy CPA zeros then meet at the binding threshold, forming a
second-order EP of the analytically continued pair
problem~\cite{Moiseyev2011,Sweeney2019}. If no direction
dominates, loss can still bind the pair without coalescence. Total momentum
changes these contributions~\cite{ZhangSong2021} and therefore selects EP sectors even on
isotropic lattices.

In this Letter, we exactly solve the lossy two-fermion Hubbard problem on
nearest-neighbor hypercubic lattices. The pair density of states fixes the loss
threshold, and an exceptional threshold is a parity-time ($\mathcal{PT}$)
symmetry transition of the pair's scattering solutions: below it two
real-energy CPA zeros are exchanged by the symmetry, increasing the loss merges
them into a second-order EP and restores the symmetry, and above it the pair is
bound and decays without oscillating. Without a dominant direction the pair
binds without the symmetry ever breaking.

\emph{Model.}---In the no-jump dynamics, two-body loss at rate $\gamma=2g$ makes
the on-site interaction purely imaginary, $U=-ig$,
\begin{equation}
\begin{aligned}
H={}&-\sum_{\bm r,\sigma}\sum_{i=1}^{d}t_i\big(c^{\dagger}_{\bm r+\bm e_i,\sigma}
c_{\bm r\sigma}+c^{\dagger}_{\bm r\sigma}c_{\bm r+\bm e_i,\sigma}\big)\\
&-ig\sum_{\bm r}n_{\bm r\uparrow}n_{\bm r\downarrow},
\end{aligned}
\label{eq:ray}
\end{equation}
where $c^{\dagger}_{\bm r\sigma}$ ($c_{\bm r\sigma}$) creates (annihilates) a
fermion of spin $\sigma$ on site $\bm r$ of a $d$-dimensional hypercubic
lattice of $N_s=L^d$ sites, $n_{\bm r\sigma}=c^{\dagger}_{\bm r\sigma}c_{\bm r\sigma}$,
and $t_i$ is the hopping along $\bm e_i$ ($t_i=t$ when isotropic), with
single-particle band $\varepsilon_{\bm k}=-2\sum_it_i\cos k_i$. In the two-particle
sector the no-jump generator is exact, a loss event simply emptying the
sector, and the one-dimensional model is
Bethe-solvable~\cite{Nakagawa2021,Wang2026}. With
$D=\sum_{\bm r}n_{\bm r\uparrow}n_{\bm r\downarrow}$ the number of doubly
occupied sites, every right eigenvector $v$, $Hv=Ev$, has
$\mathrm{Im}\,E=-g\,v^\dagger Dv/v^\dagger v\le0$: the decay rate $|\mathrm{Im}\,E|$
of a level is $g$ times its doublon weight.

On a bipartite lattice, the shift of every momentum by $\pi$ (the sublattice
sign $(-1)^{\bm r}$) composed with complex conjugation in the Bloch basis,
$\Sigma$, keeps each total-momentum sector and sends $H\to-H$: an
anti-$\mathcal{PT}$ symmetry of
$H$~\cite{GeTureci2013,PengAntiPT2016,Kawabata2019}, equivalently a
$\mathcal{PT}$ symmetry~\cite{Bender1998} of the no-jump generator $-iH$, which
we call $\mathcal{PT}$ symmetry. A symmetric level is purely imaginary and decays
without oscillating; broken levels come in pairs $E$ and $-E^*$ and decay while
oscillating. The transition concerns the solutions that feel the loss: the two
most-decaying levels on a finite lattice and, at $L\to\infty$, the CPA zeros,
which $\Sigma$ maps onto each other because it sends $E-i0$ to $-E-i0$. Since
the loss is the Hermitian part of $-iH$, increasing $g$ favors the symmetric
phase, and the EP studied below is the coalescence at which loss restores the
symmetry, a degeneracy of the kind antiunitary symmetries
constrain~\cite{Sayyad2022}.

At fixed total momentum $\bm K$ the two fermions are one particle in the
relative coordinate $\mathbf r=\mathbf r_\uparrow-\mathbf r_\downarrow$, whose
Bloch states $|\bm k\rangle$ form the pair band
$\varepsilon_2(\bm k)=\varepsilon_{\bm k}+\varepsilon_{\bm K-\bm k}$ of the
hopping term $H_0$. Up to a shift of $\bm k$ this pair kinetic energy is
$\varepsilon_2(\bm k)=-2\sum_iJ_i\cos k_i$, which defines the relative hopping
$J_i=2t_i|\cos(K_i/2)|$ along $\bm e_i$: the hops of the two fermions
interfere [Fig.~\ref{fig:channels}(a)], and $2J_i$, the half-width of the pair
band along $\bm e_i$, is the kinetic energy of the relative motion in that
direction. We order $J_1\ge J_2\ge\dots\ge J_d$; the pair density of states
is $\rho_2(E)=N_s^{-1}\sum_{\bm k}\delta[E-\varepsilon_2(\bm k)]$. The loss acts only at
$\mathbf r=0$, where the two fermions share a site: in this sector $D$ is the
projector $|d\rangle\langle d|$ onto the doublon state
$|d\rangle=b_{\bm K}^{\dagger}|0\rangle$,
$b_{\bm K}^{\dagger}=N_s^{-1/2}\sum_{\bm r}e^{i\bm K\cdot\bm r}
c^{\dagger}_{\bm r\uparrow}c^{\dagger}_{\bm r\downarrow}$, on a lattice of
$N_s$ sites. The loss is thus a rank-one perturbation of the pair band, a single
absorbing impurity in the relative motion. Every level with $\langle
d|\psi\rangle=0$ does not feel it and stays real; the levels that do are fixed
by one scalar equation, which we solve directly at $L\to\infty$. On a finite
torus, exactly one level keeps a finite decay rate as $L$ grows for $g>g_c$,
and none does for $g<g_c$; it converges to the bound pair of the infinite
lattice. Exact diagonalization of finite lattices confirms the threshold, and the
onset of the bound-pair decay rate its exceptional or non-exceptional character,
in every case studied (see Supplemental Material~\cite{SM}).

\emph{Kinetic competition.}---The scalar equation follows from the
Schr\"odinger equation $(E-H_0)|\psi\rangle=-ig|d\rangle\langle d|\psi\rangle$.
Solving it as $|\psi\rangle=-ig\langle d|\psi\rangle(E-H_0)^{-1}|d\rangle$ and
projecting back on $\langle d|$ gives the secular equation $1=-igG(E)$, with
$G(E)=\langle d|(E-H_0)^{-1}|d\rangle=N_s^{-1}\sum_{\bm k}[E-\varepsilon_2(\bm
k)]^{-1}$ the propagator of the doublon at the loss site: the amplitude that
leaves the loss site, propagates through the pair band and returns must
reproduce itself. In the infinite lattice only roots off the band, on the
physical sheet of $G$, are eigenvalues of $H$; roots on the lower lip of the
cut or on the second sheet are not. 

For incoming scattering states the equation is read on the lower lip of the
cut, $G(E-i0)$, and a real root requires $\mathrm{Re}\,G(E-i0)=0$. Summing
over $k_1$ writes $G$ as an average of one-dimensional Green's functions of
half-width $2J_1$ over the transverse energy
$W=-2\sum_{i\ge2}J_i\cos k_i$. A one-dimensional Green's function is purely
imaginary inside its band, so every term is imaginary while
$|E|+|W|<2J_1$, and
\begin{equation}
\mathrm{Re}\,G(E-i0)\equiv0\quad\text{for}\quad
|E|<2\Big(J_1-\sum_{i\ge2}J_i\Big),
\label{eq:ipv}
\end{equation}
a window that is non-empty precisely when one relative hopping exceeds the
sum of all the others (see Supplemental Material). At fixed transverse
momentum the relative motion along the dominant direction is a scattering
channel of half-width $2J_1$ shifted by $W$; an open channel contributes
only absorption to $G$, a closed one also a reactive part, and
Eq.~\eqref{eq:ipv} is the window in which every channel is open
[Fig.~\ref{fig:channels}(b)--(d)]. Inside it $G(E-i0)=i\pi\rho_2(E)$, with
$\rho_2$ the pair density of states, and the secular equation becomes the real
condition $\pi\rho_2(E)=1/g$. There $\rho_2$ is an average of shifted
one-dimensional densities of states, each even and convex inside its band, so
two real roots $\pm E_0(g)$, the CPA zeros and a $\mathcal{PT}$-broken pair,
move inward as $g$ grows and collide at the origin in a double root of the
continued secular function, a square-root branch point in $g$. Because the
loss is rank one, each root carries a single eigenvector, so the double root is
a coalescence of eigenvectors, a second-order EP, at the critical interaction
\begin{equation}
g_c=\frac{1}{\pi\rho_2(0)}=\Big[\big\langle(2|S|)^{-1}\big\rangle\Big]^{-1},
\qquad S=\sum_{i=1}^{d}J_i e^{ik_i},
\label{eq:gc}
\end{equation}
where $\langle\cdot\rangle$ averages over independent uniform phases $k_i$. The
second form holds in any dimension because $\varepsilon_2=-2\,\mathrm{Re}\,S$
is distributed as $2|S|$ times the cosine of a uniform angle independent of
$|S|$ (shifting every $k_i$ by the same angle rotates $S$), so $\rho_2$ is a
mixture of one-dimensional densities of states, the Bessel-product form of the
lattice Green function at the band center~\cite{Guttmann2010}. When no
direction dominates and the pair band has nonzero width, bipartiteness still
forces $\mathrm{Re}\,G(0)=0$. If $\rho_2(0)$ and the slope
$\mathrm{Re}\,G'(0)$ are finite, as on the isotropic cubic lattice, the
threshold is nonzero and $\mathrm{Re}\,G$ is linear in $E$ near the band
center: a single root, symmetric by itself, crosses $E=0$, and the bound pair
emerges linearly, without coalescence and without a broken pair below the
threshold.

In one dimension Eq.~\eqref{eq:gc} returns
$g_c=4t|\cos(K/2)|$~\cite{ZhangSong2021}. The continued equation reads
$E^2=g_c^2-g^2$: its real roots below $g_c$ are the broken pair, and above
$g_c$ only $-i(g^2-g_c^2)^{1/2}$ is an eigenvalue, its partner
$+i(g^2-g_c^2)^{1/2}$ lying on the second sheet; in the Bethe ansatz the EP
joins two rapidities. At $K=\pi$ the relative band is flat, the pair has
$E=-ig$, and there is no EP. On the anisotropic square lattice at $\bm K=0$
the average is an elliptic integral,
\begin{equation}
g_c(t_y)=\frac{2\pi t_x}{\ellK(t_y/t_x)} ,
\label{eq:gc2d}
\end{equation}
with $\ellK$ the complete elliptic integral of the first kind (modulus
convention), the band-center rectangular-lattice Green
function~\cite{Katsura1971}.

\begin{figure}[t]
\includegraphics[width=\columnwidth]{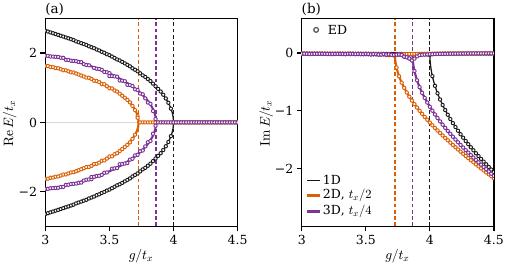}
\caption{$\mathcal{PT}$ breaking and restoration of one pair at $\bm K=0$ on
the chain, the square lattice at $t_y=t_x/2$ and the cubic lattice at
$t_y=t_z=t_x/4$: (a) $\mathrm{Re}\,E$ and (b) $\mathrm{Im}\,E$ against
$g/t_x$. Curves, $L\to\infty$: the two CPA zeros below the critical
interaction (dashed verticals) and the bound pair above it. Circles: the two
most-decaying eigenvalues of finite lattices; above the critical interaction
the second one, which stays just below $\mathrm{Im}\,E=0$, is drawn in (b)
only.}
\label{fig:roots}
\end{figure}

Figure~\ref{fig:roots} shows the transition. Below the critical interaction the
two real roots lie \emph{inside} the band, on the lower lip of the cut, so they
are not eigenvalues. They are zeros of the contact-channel $S$-matrix
$s=[1-UG(E-i0)]/[1-UG(E+i0)]$, the CPA zeros: a pair incident in the contact
channel at either energy is absorbed completely. As the loss grows they approach each other and merge at
the critical interaction, where $|s|^{2}\propto E^{4}$ and the symmetry is
restored. Above it the continued equation has two purely imaginary roots, each
symmetric by itself; only the lower one is an eigenvalue, the bound pair, which
decays without oscillating, the decaying counterpart of a repulsively bound
pair. On a finite lattice the transition is one of eigenvalues
[Fig.~\ref{fig:roots}, circles]: the two most-decaying levels form a broken
pair just below the real axis, with equal decay rates and opposite energies,
coalesce, and split along the imaginary axis into the bound pair and a slower
partner whose decay rate vanishes as $L$ grows. Complex scaling turns the CPA
zeros into eigenvalues of an auxiliary finite matrix, whose eigenvectors
coalesce at the critical interaction with the exponents of a second-order EP
(see Supplemental Material).

\begin{figure}[t]
\includegraphics[width=\columnwidth]{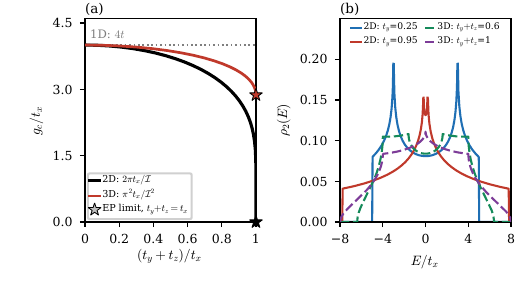}
\caption{The threshold from one dimension to three, at $\bm K=0$. (a) $g_c$
against the summed transverse hopping, in two dimensions [Eq.~\eqref{eq:gc2d},
black] and in three with $t_y=t_z$ [Eq.~\eqref{eq:gc3d}, red]; dotted, the
one-dimensional value. Stars mark $t_y+t_z=t_x$, where the window closes: $g_c$
vanishes in two dimensions and stays finite in three. (b) $\rho_2(E)$, solid in
two dimensions, dashed in three: the saddle reaching the band center makes
$\rho_2(0)$ diverge in two dimensions but is only a kink in three.}
\label{fig:dim}
\end{figure}

\emph{From one dimension to three.}---On the cubic lattice with
$t_y=t_z=t_\perp$ at $\bm K=0$ the transverse density of states is that of the
square lattice, and the average again takes a closed form,
\begin{equation}
g_c(t_\perp)=\frac{\pi^{2}t_x}{\ellK(\kappa)^{2}},\qquad
\kappa^{2}=\tfrac12\Big[1-\sqrt{1-4t_\perp^{2}/t_x^{2}}\Big],
\label{eq:gc3d}
\end{equation}
valid for $t_\perp\le t_x/2$, the band-center value of the axially anisotropic
simple-cubic Green function~\cite{DelvesJoyce2001}. New in
Eqs.~\eqref{eq:gc2d} and~\eqref{eq:gc3d} are their reading as a loss threshold
and the window of Eq.~\eqref{eq:ipv} that makes it exceptional; both reduce
to $4t_x$ as the transverse hopping vanishes [Fig.~\ref{fig:dim}(a)].

\begin{figure}[t]
\includegraphics[width=\columnwidth]{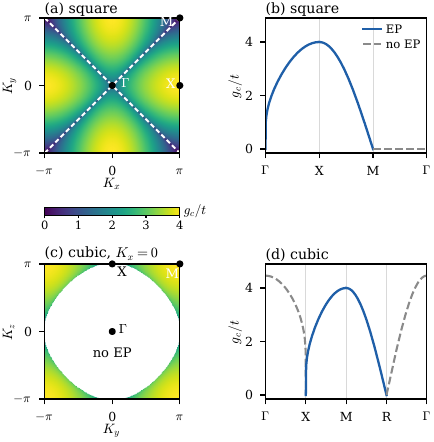}
\caption{Total momentum switches the pair EP. (a) Square lattice: $g_c$ in every
sector $\bm K$; the threshold is an EP except on the diagonals $|K_x|=|K_y|$
(dashed), where $g_c=0$. (b) $g_c$ along $\Gamma$--X--M--$\Gamma$, with
$\mathrm{X}=(\pi,0)$ and $\mathrm{M}=(\pi,\pi)$; solid, EP; dashed, no EP.
(c) Cubic lattice in the plane $K_x=0$, with $\mathrm{X}=(0,0,\pi)$ and
$\mathrm{M}=(0,\pi,\pi)$: $g_c$ in the EP sectors, white elsewhere; color scale
as in (a). (d) Cubic lattice along $\Gamma$--X--M--R--$\Gamma$, with
$\mathrm{X}=(\pi,0,0)$, $\mathrm{M}=(\pi,\pi,0)$ and $\mathrm{R}=(\pi,\pi,\pi)$;
line styles as in (b).}
\label{fig:momentum}
\end{figure}

The dimensions differ in what happens when the window closes
[Fig.~\ref{fig:dim}(b)]. Its edges are van~Hove energies of the pair band,
from the critical points at $k_1=0$, $k_{i\ge2}=\pi$ and at $k_1=\pi$,
$k_{i\ge2}=0$, and closing the window carries them to the band center. In two dimensions the saddle is a logarithmic divergence of
$\rho_2$: as $t_y\to t_x$ the critical interaction falls to zero, and the
isotropic square lattice binds a zero-momentum pair at arbitrarily weak loss.
The symmetry is then never broken. With equal hoppings the window that could
host a pair of CPA zeros has closed at $E=0$, and the
divergent density of slow pair states there lets any loss bind, so the bound
pair is purely imaginary from the weakest loss on. In three dimensions the
saddle is only a kink and $\rho_2(0)$ stays finite: the critical interaction
passes through a minimum where the window closes and rises again towards the
isotropic cubic lattice [Fig.~\ref{fig:momentum}(d), at $\Gamma$], which binds
at a finite loss. There, too, no broken
pair precedes the threshold; a single root crosses $E=0$ and the bound pair
emerges linearly. On isotropic lattices at $\bm K=0$ the threshold is a binding
transition, not a $\mathcal{PT}$ transition. The kink also limits the CPA zeros in the
exceptional regime: in three dimensions $\rho_2$ stays finite at the window
edge, so they exist only in an interval of $g$ below the critical interaction,
whereas in one and two dimensions $\rho_2$ diverges there and they persist down
to $g=0$.

Exact diagonalization shows the transition directly in the eigenvalues of
finite lattices (see Supplemental Material). Below a coupling $g_L$ the two
most-decaying levels form the broken pair just below the real axis, the role
passing from level to level as the CPA energies sweep past the discrete
spectrum. At $g_L$, which approaches the critical interaction as $L$ grows, they
coalesce; above it one descends the imaginary axis as the bound pair, while the
other rejoins the continuum. Near $g_L$ their splitting and the $c$-product
rigidity $|v\!\cdot\!v|/v^\dagger v$ of their eigenvectors vanish as
$|g-g_L|^{1/2}$, and the Petermann factor, a measure of excess noise, diverges
as $|g-g_L|^{-1}$: the eigenvectors coalesce rather than cross. Above the
threshold the bound-pair decay rate grows as $(g-g_c)^{1/2}$; on the isotropic
cubic lattice, where a finite lattice still shows a merger of its own, it grows
linearly, the signature of binding without coalescence.

\emph{Momentum sector and high dimensions.}---Because the pair at total momentum
$\bm K$ sees $J_i=2t_i|\cos(K_i/2)|$, total momentum decides whether the
threshold is a $\mathcal{PT}$ transition. On the isotropic square lattice every
sector has an EP except the diagonals $|K_x|=|K_y|$
[Fig.~\ref{fig:momentum}(a)], where $J_x=J_y$ and, as at $\bm K=0$, the
pair binds at arbitrarily weak loss without breaking the symmetry. Along
$\Gamma$--X the critical interaction is $g_c=\pi J_1/\ellK(J_2/J_1)$, the
form of Eq.~\eqref{eq:gc2d} for general $\bm K$, with $J_2/J_1=|\cos(K_x/2)|$,
and rises from zero to $4t$; along X--M the pair sees a
chain, whose band flattens at M [Fig.~\ref{fig:momentum}(b)]. On the isotropic
cubic lattice the EP sectors in the plane $K_x=0$ fill the corners
$|\cos(K_y/2)|+|\cos(K_z/2)|<1$ [Fig.~\ref{fig:momentum}(c)]. Along the zone
path of Fig.~\ref{fig:momentum}(d) the threshold is exceptional in the interior
of X--M, where the pair sees an anisotropic square lattice, and on M--R, where it
sees a chain whose band flattens at R; on $\Gamma$--X and R--$\Gamma$ the pair
binds without breaking the symmetry.

In higher dimensions one hopping must outweigh $d-1$ others, which becomes rare:
on the isotropic lattice at $\bm K=0$ no direction dominates, and
$g_c=4t/\!\int_0^\infty\!J_0(x)^{d}dx$ is never a $\mathcal{PT}$ transition and
grows as $\sqrt d$ (see Supplemental Material). Total momentum restores the EP
in every dimension: with $K_1=0$, a broken pair of CPA zeros precedes the
threshold once $\sum_{i\ge2}|\cos(K_i/2)|<1$. An isotropic lattice therefore
suffices; pairs given a finite momentum, by a moving lattice or a Bragg pulse,
could reach these sectors. Finite lattices cannot tell the two cases apart,
since every even-sided torus without an exact zero mode shows a merger; the
distinction is read from the onset of the decay rate or at $L\to\infty$.

\emph{Discussion and conclusion.}---We have exactly solved the lossy Hubbard
pair in every total-momentum sector of a nearest-neighbor hypercubic lattice
and identified when its binding threshold is exceptional. The pair density of
states at the band center sets the loss threshold, while the distribution of
relative kinetic energy among lattice directions determines whether it is an
EP, that is, whether loss restores a broken $\mathcal{PT}$ symmetry of the
pair's scattering solutions by coalescence. When one directional contribution
exceeds the others combined, two CPA zeros~\cite{Chong2010} merge at the
threshold, forming a second-order EP of the analytically continued pair
problem~\cite{Sweeney2019,Wang2021}. At zero total momentum, the isotropic
square lattice binds at arbitrarily weak loss, whereas the isotropic cubic
lattice has a finite, nonexceptional threshold; in neither does a broken pair
of CPA zeros precede the threshold. Even with isotropic hopping, pair momentum can
select sectors with an EP, in every dimension.

At the pair EP the outgoing intensity in the contact channel is suppressed
quartically near the band center; energy-resolved coherent pair scattering
could probe this signature. The exact criterion applies to an isolated pair
with a purely dissipative on-site interaction; a real on-site component breaks
the antiunitary symmetry that pins the pair spectrum. With several pairs, total
momentum no longer fixes the momentum of each pair, so pairs can bind in
sectors whose one-pair threshold is lower, and the one-pair thresholds are
dilute-limit statements. At half filling the model carries a second
antiunitary symmetry, the Shiba transformation of the down
spins~\cite{Shiba1972,Nakagawa2020} composed with conjugation, whose breaking
in the many-body ground state marks a separate EP and the subject of a companion
study~\cite{TangYangTheta}.

\begin{acknowledgments}
This work was supported by the Shenzhen Fundamental Research Program (Grant
No.~JCYJ20250604145655074), the Natural Science Foundation of Jiangsu Province (Grant
No.~BK20231320), and the Shenzhen Key Laboratory of Advanced Functional Carbon
Materials Research and Comprehensive Application (Grant
No.~ZDSYS20220527171407017).
\end{acknowledgments}

\textit{Data availability.}---The data that support the findings of this Letter
are contained in the paper and its Supplemental Material; additional data and
source code are available from the corresponding authors upon reasonable
request.

\end{document}


\title{Supplemental Material for ``Kinetic competition controls loss-induced
exceptional points of Hubbard pairs''}

\author{Ho-Kin Tang\textsuperscript{1,3,}}
\email[Corresponding author: ]{denghaojian@hit.edu.cn}
\author{Xiaosen Yang\textsuperscript{2,}}
\email[Corresponding author: ]{yangxs@ujs.edu.cn}
\affiliation{\textsuperscript{1}School of Science, Harbin Institute of Technology, Shenzhen 518055, China\\
\textsuperscript{2}Department of Physics, Jiangsu University, Zhenjiang 212013, China\\
\textsuperscript{3}Shenzhen Key Laboratory of Advanced Functional Carbon Materials
Research and Comprehensive Application, Shenzhen 518055, China}

\date{\today}

\maketitle

In this Supplemental Material we give the numerical verification and the
derivations behind the Letter. Sec.~\ref{sec:ed} measures the one-pair
threshold by exact diagonalization alone, without the rank-one equation: the
finite-lattice exceptional point approaches $g_c$, and the decay rate of the
bound pair has a square-root onset where the rule predicts an exceptional point
and a linear one where it predicts none. It then follows the $\mathcal{PT}$
transition through the spectrum of finite lattices, describes the numerical
method, and tests the coalescence of eigenvectors on finite lattices and
directly in the continuum. Sec.~\ref{sec:derive} derives the
exceptional point from the symmetry and the rank-one reduction, in one, two,
three and higher dimensions, through an arcsine-mixture representation of the
pair density of states (Sec.~\ref{sec:am}), and shows that on a torus exactly one level of a
fixed-momentum pair sector keeps a finite decay rate as the lattice grows, the
bound pair, and only when $g>g_c$. Sec.~\ref{sec:1d:ba} derives the
one-dimensional case again by Bethe ansatz. Throughout, $\ellK(k)$ is the
complete elliptic integral of the first kind in the modulus convention, since
$K$ denotes total momentum and $\mathcal{K}$ complex conjugation; the window of
real energies on which the real part of the pair Green function vanishes
identically, Eq.~(2) of the Letter, is called the open-channel window, or
simply the window; and the antiunitary symmetry $\Sigma$ is the $\mathcal{PT}$
symmetry of the Letter.

\section{Exact diagonalization and numerical verification}
\label{sec:ed}

\subsection{The one-pair threshold}
\label{sec:edthr}

The threshold $g_c=1/[\pi\rho_2(0)]$ and the bound-state rate $y^*(g)$ follow from the
rank-one secular equation. To test them without it we diagonalize the matrix of
Eq.~(1) of the Letter directly, for one pair at total momentum $\bm K$ in the
relative coordinate, using only lattice symmetry: translations fix $\bm K$, and the
reflections and permutations of equivalent directions that leave the loss site
fixed reduce the matrix to its symmetric sector. The reduction is exact, since a
state orthogonal to that sector has no amplitude where the two particles meet and
its eigenvalue is real and independent of $g$; it reproduces every decaying
eigenvalue of the unreduced matrix to $10^{-13}$. Eigenvalues come from dense
diagonalization or from Arnoldi iteration for the most-decaying level, and no step
uses $G(E)$ or the continuum. The sectors have effective hoppings
$\tau=(1,a,\dots,a)$ with $a=1/[2(d-1)]$, where the rule predicts an EP in every
dimension, and the isotropic sectors at $\bm K=0$, where it predicts none.

\begin{figure}[t]
\includegraphics[width=\textwidth]{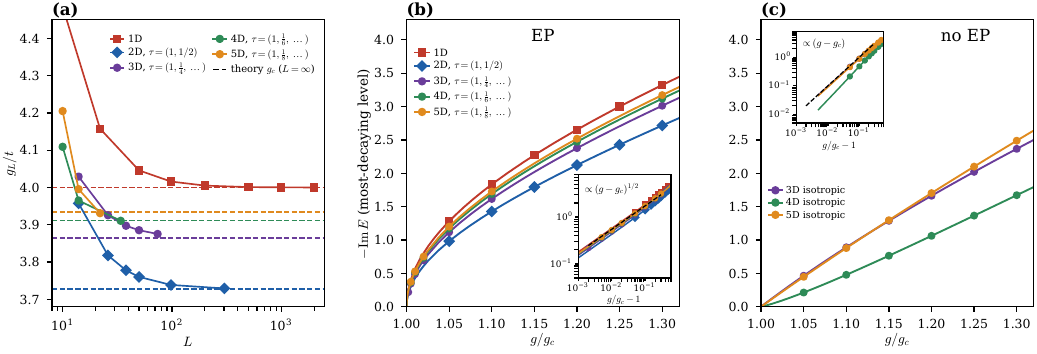}
\caption{(Color online) The threshold from exact diagonalization alone. Symbols:
eigenvalues of Eq.~(1) of the Letter; lines: rank-one theory at $L=\infty$, for
comparison only. (a) Finite-lattice exceptional point $g_L$, at which the
most-decaying level reaches the imaginary axis, against $L$, in the exceptional
sectors $\tau=(1,a,\dots,a)$, $a=1/[2(d-1)]$; dashed, $g_c$. For $d=2$, $3$ only
lattices with $L\equiv2\pmod 3$ are shown (the other residue class converges to
the same $g_c$ from the other side of the exact $E=0$ level at
$\bm q=(\pi/3,\pi,\dots,\pi)$); the $d=4$ points at $L=\EDomitFourA$ and
$\EDomitFourB$, which undershoot $g_c$ by one to two percent, are not shown.
(b) Exceptional sectors: decay rate $y=-\mathrm{Im}\,E$ of the most-decaying level
at the largest $L$, converged in $L$, against $g/g_c$; inset, log-log, slope $1/2$.
(c) Isotropic sectors at $\bm K=0$ (no EP); inset, slope $1$ (up to a
logarithm at $d=4$).}
\label{fig:ed}
\end{figure}

Figure~\ref{fig:ed} collects three measurements. (a) On a finite lattice the two
most-decaying levels form a $\Sigma$ pair below a coupling $g_L$ and leave it as
two levels on the imaginary axis above; $g_L$, located by bisection on the
eigenvalues alone, is an exceptional point of the physical Hamiltonian, and it
approaches $g_c$ as $L$ grows, up to $L=\EDLringOneD$ on the chain and
$\EDLmergeTwo$, $\EDLmergeThree$, $\EDLmergeFour$, $\EDLmergeFive$ for $d=2$--$5$.
(b) In the exceptional sectors the decay rate of the bound pair, converged in $L$,
rises as $y=A(g-g_c)^{\nu}$; a fit with $A$, $g_c$ and $\nu$ free gives
$\nu=\EDnuEPmin$--$\EDnuEPmax$ for $d=3$--$5$ and recovers $g_c$ to
$\EDgcEPdev\%$, the square-root onset of a branch point. (c) In the isotropic
sectors, at $L$ up to $\EDLonsetThree$, $\EDLonsetFour$ and $\EDLonsetFive$ for
$d=3$, $4$, $5$, the same fit gives $\nu=\EDnuISOmin$--$\EDnuISOmax$, the linear
onset of a threshold without an EP (at $d=4$ the slope carries a logarithm,
Sec.~\ref{sec:hd}, which the fit absorbs into $\nu$). In both cases the rates sit on $y^*(g)$ of the
rank-one theory.

\subsection{The $\mathcal{PT}$ transition on finite lattices}
\label{sec:pt}

\begin{figure*}[t]
\includegraphics[width=\textwidth]{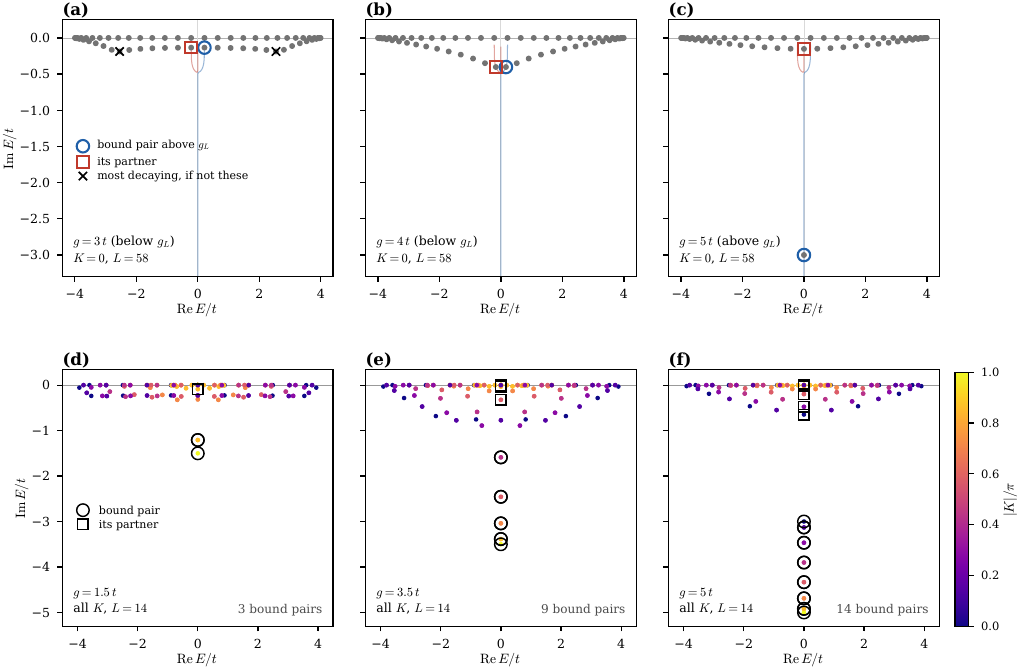}
\caption{(Color online) The $\mathcal{PT}$ transition in the spectrum of the
chain, from exact diagonalization. (a)--(c) Every level of the $\bm K=0$ sector,
$L=\SPLtrack$, at $g=\SPgTa\,t$, $\SPgTb\,t$ and $\SPgTc\,t$, with
$g_L=\SPgLtrack\,t$. Circle and square: the two levels that coalesce at $g_L$,
the circle becoming the bound pair above $g_L$; thin lines, their paths over the
whole scan; crosses, the two most-decaying levels where they are not these.
(d)--(f) All $L^2$ levels at $L=\SPLsmall$ and $g=\SPgA\,t$, $\SPgB\,t$ and
$\SPgC\,t$, coloured by $|K|/\pi$. Rings: the bound pair of each sector that has
one, the most-decaying level whose doublon weight is unchanged when $L$ is
doubled; squares, its partner.}
\label{fig:pt}
\end{figure*}

Figure~\ref{fig:pt} shows the transition in the full spectrum of the chain,
from exact diagonalization alone. Panels (a)--(c) follow every level of the
$\bm K=0$ sector of a ring of $L=\SPLtrack$ sites through the transition, at
$g=\SPgTa\,t$, $\SPgTb\,t$ and $\SPgTc\,t$; on this ring the two levels that
coalesce do so at $g_L=\SPgLtrack\,t$, close to $g_c=\gcOneD\,t$. At $g=0$ every
level is real. As the loss grows the levels with weight on the doublon acquire
decay rates and bend into the lower half-plane, while the levels without that
weight stay on the real axis. Below $g_L$ the two most-decaying levels form a
$\Sigma$-broken pair, $E$ and $-E^{*}$, at opposite energies with equal decay
rates [crosses in (a)]. The role is handed from level to level as the CPA
energies $\pm E_0(g)$ of the infinite lattice sweep past the discrete spectrum,
so the pair that finally coalesces (circle and square) is the one nearest the
band centre. As $g$ approaches $g_L$ it moves inward and down [panel (b)]; at
$g_L$ its two members meet at $\mathrm{Re}\,E=0$; above $g_L$ they split along
the imaginary axis [panel (c)]. One, the bound pair, descends to
$-iy^{*}(g)$ of the infinite lattice, while the other, the slower partner, rises
back towards the real axis and rejoins the continuum: its doublon weight and
decay rate fall as $1/L$ (Sec.~\ref{sec:sp}).

Panels (d)--(f) show the same transition in every momentum sector at once, all
$L^2$ levels of a ring of $L=\SPLsmall$ sites at $g=\SPgA\,t$, $\SPgB\,t$ and
$\SPgC\,t$. Each sector has its own threshold, $4t|\cos(K/2)|$ in one dimension,
so the sectors bind one after another as $g$ grows, starting near $K=\pi$
where the relative band is narrowest. At most one level per sector binds; the
bound pair is identified without the secular equation, as the most-decaying
level of its sector whose doublon weight does not change when $L$ is doubled,
and its partner stays near the real axis. Whether the meeting at $g_L$ is a
coalescence of eigenvectors rather than a crossing is tested in
Sec.~\ref{sec:diag}.

\subsection{Method}
\label{sec:method}

\paragraph{Momentum-projected exact diagonalisation.}
We work in the plane-wave basis at fixed total momentum. The interaction
$D=(1/N_s)\sum_q\rho^\uparrow_{q}\rho^\downarrow_{-q}$ retains the spin
factorisation, so a sector is a Kronecker structure on the two spin species and
the matrix--vector product costs one pass over $q$. The sector dimension is
$\sim\binom{N_s}{n_\uparrow}\binom{N_s}{n_\downarrow}/N_s$, which at fixed
particle number is \emph{polynomial} in $N_s$. Configurations are stored as
sorted index lists with combinadic ranking. Reflection halves the work, since $K$ and
$N_s-K$ have identical spectra, so scanning $K=0\ldots N_s/2$ is complete. In
the two-particle sector the interaction is exactly rank one at fixed total
momentum (Sec.~\ref{sec:derive:prelim}); the secular problem is then
diagonal-plus-rank-one and is solved in $O(L)$ operations with a closed-form
inner momentum sum, and the continuum results of the Letter are obtained
directly at $L\to\infty$ by quadrature of the same secular function.

The construction is validated by requiring that the union over all momentum
sectors reproduce the real-space full-sector spectrum, eigenvalue for
eigenvalue, at generic complex $U$. That single test exercises the basis, the
momentum bookkeeping, the Jordan--Wigner signs in $\rho_q$, the $1/N_s$
normalisation and the boundary conditions simultaneously; it passes at
$\lesssim10^{-13}$ in one and two dimensions, for both periodic and antiperiodic
boundaries. Two further gates run throughout: at $U=0$ the spectrum reproduces
the sums of single-particle energies exactly, and at real $U$ it is real. The
$O(L)$ secular solver is gated against dense diagonalisation to
$\sim10^{-14}$. The one-pair results of Sec.~\ref{sec:edthr}, of
Fig.~\ref{fig:diagnostics} and of the finite-lattice points of Fig.~2 of the
Letter use the relative-coordinate, symmetry-reduced matrix described in
Sec.~\ref{sec:edthr}, diagonalized densely or, for the largest sectors, by
Arnoldi iteration for the most-decaying levels; Fig.~2 of the Letter uses
$L=\FtwoLchain$ for the chain, up to $\FtwoLsquareMax$ for the square lattice
and up to $\FtwoLcubicMax$ for the cubic lattice. The full spectra of
Fig.~\ref{fig:pt}(d)--(f) use the two-particle momentum basis above.

\paragraph{Comparing spectra.}
Because $\Sigma$ makes the extremal level a degenerate $\pm\mathrm{Re}\,E$
pair, two solvers may return different members of that pair. Comparisons
between solvers are therefore made by optimal assignment between the two
eigenvalue sets rather than by sorting on any single component, and every solve
is checked against the rigorous bound $\mathrm{Im}\,E\in[-gD_{\max},0]$.

\subsection{Eigenvector coalescence, on finite lattices and in the continuum}
\label{sec:diag}

\begin{figure*}[t]
\includegraphics[width=0.86\textwidth]{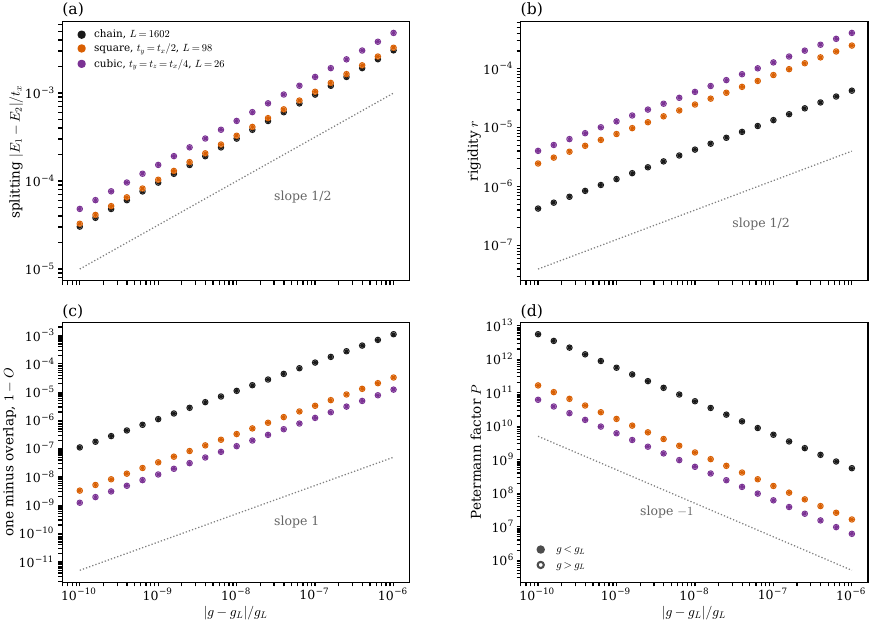}
\caption{(Color online) Eigenvector coalescence at the finite-lattice
exceptional point $g_L$, from exact diagonalization of Eq.~(1) of the Letter, one
pair at $\bm K=0$, for the chain ($L=\CoalLOne$), the square lattice at
$t_y=t_x/2$ ($L=\CoalLTwo$) and the cubic lattice at $t_y=t_z=t_x/4$
($L=\CoalLThree$): (a) splitting $|E_1-E_2|$ of the two coalescing levels,
(b) $c$-product rigidity $r=\sqrt{r_1r_2}$ of their eigenvectors, (c) one minus
their Hermitian overlap $O$, (d) Petermann factor $P=1/(r_1r_2)$, against
$|g-g_L|/g_L$; filled below $g_L$, open above. Dotted guides: slopes $1/2$,
$1/2$, $1$ and $-1$.}
\label{fig:diagnostics}
\end{figure*}

On a finite lattice the meeting of Fig.~\ref{fig:pt} is a coalescence of
eigenvectors, not a level crossing (Fig.~\ref{fig:diagnostics}), shown on the
largest lattices we diagonalized, the chain at $L=\CoalLOne$, the square
lattice at $t_y=t_x/2$ with $L=\CoalLTwo$ and the cubic lattice at
$t_y=t_z=t_x/4$ with $L=\CoalLThree$. Approaching $g_L$ from either side, the
splitting $|E_1-E_2|$ of the two levels vanishes as $|g-g_L|^{1/2}$ [panel (a)].
The $c$-product rigidity $r$ of their eigenvectors vanishes in the same way
[panel (b)], and one minus their Hermitian overlap $O$ as $|g-g_L|$
[panel (c)], with fitted exponents $\CoalRig$ and $\CoalOvl$ in every case and on both
sides. The Petermann factor $P=1/(r_1r_2)$, the excess noise of the
pair of modes, diverges as $|g-g_L|^{-1}$ [panel (d)]. At a crossing the eigenvectors would
stay distinct, and $r$ and $P$ would stay finite. The same holds in one, two
and three dimensions, and $g_L$ approaches $g_c$ as $L$ grows
[Fig.~\ref{fig:ed}(a)]. The window in $g$ in which the coalescing pair is
isolated from its neighbours shrinks as $1/L^{2}$, because the pair levels
near the band centre are spaced by $\propto1/L$ while the CPA energies move as
$|g-g_c|^{1/2}$; the panels stay inside it for every lattice. Beyond it the
most-decaying pair is handed to the next levels, and at
$L=\CoalLOne$ the splitting there already takes the infinite-lattice value
$2|E_0(g)|$. The prefactors of the four laws inside the window depend on $L$; the
exponents do not.

The two-body $\Sigma$ exceptional point is established analytically, by the
double root of the secular function $F(z)=1+igG(z)$ at $E=0$ and $g=g_c$
(Sec.~\ref{sec:derive}). The diagnostics of this subsection reconfirm that exact
result with the same eigenvector tests, read directly at $L\to\infty$, with $g_c$ taken from the closed forms and
no fitted threshold. Below $g_c$ the two roots lie inside the pair band, on the
lower lip of the cut, and are zeros of the pair $S$-matrix rather than
eigenvalues [Eq.~\eqref{eq:1d:sratio}]. To give them eigenvectors we use complex
scaling in the sense of Aguilar, Balslev, Combes and Simon, in the form used
throughout non-Hermitian quantum mechanics~\cite{S_Moiseyev2011}: the momentum
contour of the pair sum is deformed into the complex plane so that the pair
continuum is lifted into the \emph{upper} half-plane, which exposes the
lower-lip roots as ordinary, isolated eigenvalues of a finite complex-symmetric
matrix. The roots and the derivative $F'$ at them do not depend on the
deformation; the exposing contour itself is not unique.

On that matrix we read, for the coalescing pair $v_\pm$, the splitting
$|E_+-E_-|$; the $c$-product rigidity
$r_\pm=|v_\pm\!\cdot\!v_\pm|/(v_\pm^\dagger v_\pm)$ of each member, combined as
$r=\sqrt{r_+r_-}$; the Hermitian overlap $O=|\langle v_+|v_-\rangle|$ of the
normalised pair; and $P=1/(r_+r_-)$. For a complex-symmetric matrix the
per-mode Petermann factor of an eigenvector is $1/r^{2}$, so $P$ is the
geometric mean of the per-mode Petermann factors $1/r_+^{2}$ and $1/r_-^{2}$ of
the coalescing pair, and it diverges as $|g-g_c|^{-1}$ because each $r_\pm$
vanishes as $|g-g_c|^{1/2}$. For the same reason the rigidity and the overlap
are two faces of one self-orthogonality, and they test the same coalescence in
two normalisations rather than independently.

On the chain and the anisotropic square lattice, $t_y/t_x=\csTyList$, the splitting and
$F'$ follow the exponents $\csAlphaLo$--$\csAlphaHi$ against the exact $1/2$ on
both sides of $g_c$, over $\csFitLo\le|g-g_c|/g_c\le\csFitHi$, and their
amplitudes equal the closed-form prefactors to a relative deviation of at most
$\csAmpDevMax$; the rigidity and one minus the overlap follow the exponents
$\csRigAlphaLo$--$\csRigAlphaHi$ and $\csOvlAlphaLo$--$\csOvlAlphaHi$ against
$1/2$ and $1$. The cubic lattice with $t_y=t_z$ inside the window gives the same exponents. The values of $r$, $O$ and $P$ are
properties of the complex-scaled matrix rather than of the continuum problem:
they depend on the contour, by up to tens of percent between the contours used,
whereas their exponents, and the roots and $F'$ themselves, do not, and only the
latter enter any gate. The two-body problem is therefore the one case in this
Supplement where the eigenvector diagnostics are a confirmation rather than the
evidence.

\clearpage
\section{The two-body exceptional point, from one dimension to any}
\label{sec:derive}

\subsection{The model, its antiunitary symmetry, and the rank-one reduction}
\label{sec:derive:prelim}

On a $d$-dimensional hypercubic lattice with nearest-neighbour hoppings $t_i>0$
the Hamiltonian on the dissipative ray, $U=-ig$ with $g>0$, is Eq.~(1) of the
Letter,
\begin{equation}
H=-\sum_{\bm r,\sigma}\sum_{i=1}^{d}t_i\big(c^{\dagger}_{\bm r+\bm e_i,\sigma}
c_{\bm r\sigma}+c^{\dagger}_{\bm r\sigma}c_{\bm r+\bm e_i,\sigma}\big)
-ig\sum_{\bm r}n_{\bm r\uparrow}n_{\bm r\downarrow}
\equiv H_0-igD ,
\label{eq:sm:ray}
\end{equation}
where $c^{\dagger}_{\bm r\sigma}$ creates a fermion of spin $\sigma$ at site
$\bm r$, $n_{\bm r\sigma}=c^{\dagger}_{\bm r\sigma}c_{\bm r\sigma}$, $t_i$ is the
hopping along the unit vector $\bm e_i$, $H_0$ is the hopping term and
$D=\sum_{\bm r}n_{\bm r\uparrow}n_{\bm r\downarrow}$ the doublon number; the
symbol $K$ is reserved for the total momentum. $H_0$ and $D$ are both real in
the site basis, so $H$ is complex \emph{symmetric},
$H^{T}=H$: left eigenvectors are transposes of right ones and a single vector
carries both. Two products therefore appear, and we say for each quantity
which one it uses. The $c$-product
$\langle O\rangle_c=(v\!\cdot\!Ov)/(v\!\cdot\!v)$, taken without conjugation,
is the biorthogonal expectation value of a complex-symmetric
matrix~\cite{S_Moiseyev2011}; it is stationary under variations
of $v$, and it is the product in which the rigidity of Sec.~\ref{sec:ed} is
defined. It is
complex in general, and $\langle D\rangle_c$ is \emph{not} $-\mathrm{Im}\,E/g$.
The Hermitian (sesquilinear) form gives an exact identity instead: for any
right eigenvector, $v^\dagger Hv=E\,v^\dagger v$ with $v^\dagger H_0v$ real and
$v^\dagger Dv\ge0$, so
\begin{equation}
\mathrm{Im}\,E=-g\,\frac{v^\dagger Dv}{v^\dagger v}\le0
\label{eq:sm:imE}
\end{equation}
for every eigenvalue on the ray, at an exceptional point included. The whole
spectrum lies in the strip $-gD_{\max}\le\mathrm{Im}\,E\le0$, and no level on
the ray, on any lattice, has $\mathrm{Im}\,E>0$. Equation~\eqref{eq:sm:imE}
holds on the ray only, where $v^\dagger Uv$ is purely imaginary. Wherever this
Supplement writes $\langle D\rangle$ without a subscript, the Hermitian
expectation of Eq.~\eqref{eq:sm:imE} is meant, so that
$\langle D\rangle=-\mathrm{Im}\,E/g$ exactly.

\paragraph{$\Sigma$: the momentum shift by $\bm\pi$ composed with conjugation.}
Let $P$ be the unitary shift of every single-particle momentum by
$\bm\pi=(\pi,\dots,\pi)$, $c_{\bm k\sigma}\to c_{\bm k+\bm\pi,\sigma}$, let
$\mathcal{K}_{\bm k}$ be complex conjugation in the \emph{momentum} basis, which
keeps every $\bm k$ and hence every total momentum $\bm K$, and set
$\Sigma=P\circ\mathcal{K}_{\bm k}$. Conjugation in the site basis maps plane waves $e^{i\bm k\cdot\bm r}$ to $e^{-i\bm k\cdot\bm r}$ and
the sector $\bm K$ to $-\bm K$; the two conjugations differ by the spatial
inversion $\bm r\to-\bm r$, a unitary symmetry of $H$, so either choice gives
the symmetry of the spectrum derived below, but only $\mathcal{K}_{\bm k}$ acts
within a fixed-$\bm K$ sector. The shift sends
$\varepsilon_{\bm k}\to\varepsilon_{\bm k+\bm\pi}=-\varepsilon_{\bm k}$, so
the hopping term changes sign, $H_0\to-H_0$, while the density operators
$\rho_{\bm q}=\sum_{\bm k}c^\dagger_{\bm k+\bm q}c_{\bm k}$ are untouched because
both operators shift together; hence $D\to D$. The unitary shift alone
therefore gives $PHP^{-1}=-H_0-igD=-H^{*}$, and composing it with conjugation,
which reverses the sign of $-ig$, makes $\Sigma$ \emph{anticommute} with $H$:
\begin{equation}
\Sigma H\Sigma^{-1}=PH^{*}P^{-1}=-H_0+igD=-H .
\label{eq:sm:sigma}
\end{equation}
For $Hv=Ev$ this gives $H(\Sigma v)=-\Sigma Hv=-E^{*}\,\Sigma v$, so
$E\to-E^{*}$: the spectrum of a fixed-momentum sector is symmetric about the
\emph{imaginary} axis, and a level mapped to itself has $\mathrm{Re}\,E=0$. An
antiunitary that anticommutes with $H$ is an anti-$\mathcal{PT}$ symmetry; in
the classification of Ref.~\cite{S_Kawabata2019} it is of particle--hole type,
PHS$^\dagger$-type. Equivalently, $\Sigma$ commutes with the generator $-iH$ of
the no-jump evolution, of which it is a $\mathcal{PT}$ symmetry; the Letter
uses that name. The construction needs
$\varepsilon_{\bm k+\bm\pi}=-\varepsilon_{\bm k}$, that is a bipartite
nearest-neighbour band: a next-nearest-neighbour hopping $t'$ is even under the
shift and destroys $\Sigma$. The shift moves the total momentum by $N\bm\pi$ modulo $2\pi$,
so the sector $\bm K$ maps to itself precisely when the total particle number
$N$ is even; $\Sigma$ imposes no condition on the filling. Equation~\eqref{eq:sm:sigma} holds only at
purely imaginary $U$, since the interaction must be purely imaginary for conjugation
to reverse it; with $U=U_0-ig$ and $U_0\neq0$, $\Sigma$ is not a symmetry,
because $U_0D$ is not reversed by conjugation. The symmetry is a property of
the dissipative ray itself.

\paragraph{The pair propagator and the rank-one reduction.}
Let $d_{\bm r}^{\dagger}=c^{\dagger}_{\bm r\uparrow}c^{\dagger}_{\bm r\downarrow}$
create a doublon on site $\bm r$, and
$b_{\bm K}^{\dagger}=N_s^{-1/2}\sum_{\bm r}e^{i\bm K\cdot\bm r}d_{\bm r}^{\dagger}
=N_s^{-1/2}\sum_{\bm k}c^{\dagger}_{\bm k\uparrow}c^{\dagger}_{\bm K-\bm k,\downarrow}$
a doublon Bloch wave of total momentum $\bm K$. In the sector of one up and one
down particle $n_{\bm r\uparrow}n_{\bm r\downarrow}=d_{\bm r}^{\dagger}d_{\bm r}$,
so $D=\sum_{\bm K}b_{\bm K}^{\dagger}b_{\bm K}$, and within the sector of fixed
total momentum the interaction is the single projector
$U\,b_{\bm K}^{\dagger}b_{\bm K}$: it couples every relative momentum to every
other with the same amplitude $1/N_s$, and the problem is exactly rank one. It
is also confined to the spin singlet. The on-site triplet pair operators vanish
by the Pauli principle, so $d_{\bm r}^{\dagger}$ creates an on-site singlet; the
$S_z=0$ triplet combinations are annihilated by $b_{\bm K}$ and keep their
free, real energies; and $H_0$ and $D$ both commute with the total spin, so
this separation holds for every complex $U$. We define the pair Green function
as the propagator of the doublon under the kinetic term alone,
\begin{equation}
G_{\bm K}(E)=\langle0|\,b_{\bm K}\,(E-H_0)^{-1}\,b_{\bm K}^{\dagger}\,|0\rangle
=\frac{1}{N_s}\sum_{\bm k}\frac{1}{E-\varepsilon_2(\bm k)},\qquad
\varepsilon_2(\bm k)=\varepsilon_{\bm k}+\varepsilon_{\bm K-\bm k} ,
\label{eq:sm:Gop}
\end{equation}
the resolvent of $H_0$ in the doublon channel, whose boundary values
$G(E\pm i0)$ are the retarded and advanced propagators. For an eigenstate of
$H=H_0+U\,b_{\bm K}^{\dagger}b_{\bm K}$ at energy $E$ off the free spectrum,
projecting $(E-H_0)|\psi\rangle=U\,b_{\bm K}^{\dagger}|0\rangle\langle0|b_{\bm K}|\psi\rangle$
onto $\langle0|b_{\bm K}(E-H_0)^{-1}$ gives the Koster--Slater condition
$1=U\,G_{\bm K}(E)$, which on the ray reads $1=-ig\,G_{\bm K}(E)$. In the
thermodynamic limit the free spectrum becomes a branch cut of $G$. For a real
energy inside it, $1=-igG(E-i0)$ imposes the incoming-wave boundary condition
for a zero of the contact-channel $S$-matrix, a CPA zero; such a root is not a
normalizable eigenvalue of $H$. A bound-pair eigenvalue instead solves the
equation on the physical sheet below the real axis, consistently with
Eq.~\eqref{eq:sm:imE}. On the upper lip the corresponding zero requires gain,
$U=+ig$.

Nothing in this construction is specific to a dimension. The sections that
follow are organised by what changes, which is only the pair band
$\varepsilon_2$ and hence the pair density of states. The last of them,
Sec.~\ref{sec:sp}, returns to finite lattices and shows that the
$L\to\infty$ analysis describes the limit of their spectra. Their conventions are
fixed in Sec.~\ref{sec:1d}: the pair band at total momentum $\bm K$ is that of
$\bm K=0$ with the effective hoppings $\tau_i=t_i|\cos(K_i/2)|$ (the Letter writes
the relative hopping $J_i=2\tau_i$, defined by the pair kinetic energy
$\varepsilon_2=-2\sum_iJ_i\cos k_i$, so that $4\tau_i=2J_i$ is the kinetic energy
of the relative motion along $\bm e_i$), after which
$\bm K=0$ is taken and the hoppings are ordered $t_1\ge t_2\ge\dots\ge t_d>0$ (the Letter's
$S=\sum_iJ_ie^{ik_i}$ is twice the $S=\sum_it_ie^{ik_i}$ used below);
the secular function is $F(z)=1-UG(z)=1+igG(z)$; the one-dimensional Green
function of a cosine band of half-width $a$ is written
$G_a(z)=(z-a)^{-1/2}(z+a)^{-1/2}$; and $\Sigma$ denotes the antiunitary
symmetry $E\mapsto-E^{*}$ of Eq.~\eqref{eq:sm:sigma}. Numerical values quoted
in brackets in these sections come from numerical checks of the closed forms
and from the figure captions; nothing in the derivations depends on them.

The special functions in which $g_c$ is expressed below are not new, and it
is worth saying at the outset what is taken as known and what is claimed.
Since $g_c=1/\pi\rho_2(0)$ and $\rho_2$ is the density of states of the pair
band, each closed form is the band-centre value of a known lattice Green
function: the complete elliptic integral of the rectangular lattice,
Eq.~\eqref{eq:2d:gc}, is the band-centre value of the anisotropic
square-lattice Green function~\cite{S_Katsura1971}; the
$\ellK(k)^{2}$ form of the cubic lattice with two equal hoppings,
Eq.~\eqref{eq:3d:gc}, is that of the anisotropic simple-cubic Green
function~\cite{S_Delves2001}; the $\Gamma(1/3)$ form of the
isotropic cubic value, Eq.~\eqref{eq:3d:iso}, belongs to the Watson-integral
literature; and the Bessel-product (Laplace)
representation of Eq.~\eqref{eq:am:watson} is the standard time-domain form of
the hypercubic Green function~\cite{S_Guttmann2010}. These
evaluations are taken as known. What is claimed here is the physics that
selects them: the identification of the loss threshold with
$g_c=1/\pi\rho_2(0)$, the statement that the threshold is an exceptional point
exactly when $\mathrm{Re}\,G(E-i0)$ vanishes on an interval, the open-channel window, the
characterisation of that interval through the arcsine-mixture representation
of $\rho_2$ (Sec.~\ref{sec:am}), and the kinetic-competition criterion (the polygon rule) that follows from it. In
one dimension, where the threshold is $\gcOneD\,t$, the threshold, the real
sub-threshold roots and their reading as complete pair annihilation are published by Song and
co-workers~\cite{S_LiZhangSong2014,S_ZhangSong2021}, and
Sec.~\ref{sec:1d} reproduces them as the $d=1$ case of the general argument.

\begingroup\let\section\subsection \let\subsection\subsubsection
\section{One dimension}
\label{sec:1d}

The one-dimensional case is derived here by the resolvent (Koster--Slater) route,
the one that generalises to any dimension; the Bethe ansatz of
Sec.~\ref{sec:1d:ba} derives it a second time and agrees on every statement.
On a ring of $L$ sites,
\begin{equation}
H=-t\sum_{j\sigma}\big(c^\dagger_{j+1\sigma}c_{j\sigma}+\mathrm{h.c.}\big)
+U\sum_j n_{j\uparrow}n_{j\downarrow},\qquad U=-ig,\quad g>0 ,
\label{eq:1d:H}
\end{equation}
in the sector of one $\uparrow$ and one $\downarrow$ particle with total
momentum $K$. The one-particle dispersion is $\varepsilon(k)=-2t\cos k$, and the
pair band is
\begin{equation}
\varepsilon_2(k)=\varepsilon(k)+\varepsilon(K-k)
=-2t\big[\cos k+\cos(K-k)\big]
=-4t\cos(K/2)\,\cos q,\qquad q=k-\tfrac K2 ,
\label{eq:1d:band}
\end{equation}
using $\cos A+\cos B=2\cos\frac{A+B}2\cos\frac{A-B}2$. It is a single cosine of
half-width
\begin{equation}
\boxed{\;a\equiv a(K)=4t\,|\cos(K/2)|\;}
\label{eq:1d:a}
\end{equation}
The absolute value is not physical input: $K$ is defined modulo $2\pi$, so
$\cos(K/2)$ may come out with either sign, and the shift $q\to q+\pi$ flips it
while leaving the set of energies unchanged. Everything below is at fixed $K$,
with $a>0$; the flat case $K=\pi$ ($a=0$) is the $\eta$-pairing sector, treated
separately at the end of this section.

\paragraph{The interaction is rank one.}
Let $b_K^\dagger=L^{-1/2}\sum_je^{iKr_j}c^\dagger_{j\uparrow}c^\dagger_{j\downarrow}$
and $|\beta\rangle=b^\dagger_K|0\rangle$. In this sector every matrix element of
$\sum_jn_{j\uparrow}n_{j\downarrow}$ between two basis states of the same total
momentum equals $1/L$, so the interaction is $U|\beta\rangle\langle\beta|$.
For an eigenvector with $E\notin\mathrm{spec}\,H_0$, projecting
$(E-H_0)|\psi\rangle=U|\beta\rangle\langle\beta|\psi\rangle$ onto
$\langle\beta|$ gives the Koster--Slater condition
\begin{equation}
1=U\,G(E),\qquad G(z)=\frac1L\sum_q\frac1{z-\varepsilon_2(q)} ,
\qquad \psi(q)\propto\frac1{E-\varepsilon_2(q)} .
\label{eq:1d:KS}
\end{equation}
Since the anti-Hermitian part of $H$ is $-ig\,|\beta\rangle\langle\beta|\le0$,
every eigenvalue obeys $\mathrm{Im}\, E\le0$, so the physical boundary value on the band
is the lower lip $G(E-i0)$.

\paragraph{Density of states and Green function.}
With $\varepsilon=-a\cos q$ and $q$ uniform, $|d\varepsilon/dq|=a\sin q=\sqrt{a^2-\varepsilon^2}$, so
\begin{equation}
\rho_a(\varepsilon)=\frac1{\pi\sqrt{a^2-\varepsilon^2}},\qquad|\varepsilon|<a
\qquad\text{(arcsine law)} ,
\label{eq:1d:arcsine}
\end{equation}
and its Stieltjes transform is, by the substitution $u=e^{iq}$ and one residue,
\begin{equation}
G_a(z)=\int\frac{\rho_a(\varepsilon)\,d\varepsilon}{z-\varepsilon}
=\frac1{(z-a)^{1/2}(z+a)^{1/2}} ,
\label{eq:1d:G}
\end{equation}
with principal roots in each factor. This is the branch analytic off $[-a,a]$
with $G_a\simeq1/z$ at infinity; it is \emph{not} the principal root of
$z^2-a^2$, which would place a spurious cut on the imaginary axis and reverse
the sign of $G_a(-iy)$.

\paragraph{Boundary values.} From $G_a^2=1/(z^2-a^2)$, continuity, and the
asymptotics alone (no Sokhotski--Plemelj needed),
\begin{equation}
G_a(-iy)=\frac{i}{\sqrt{y^2+a^2}},\qquad
G_a(E-i0)=\begin{cases}\dfrac{i}{\sqrt{a^2-E^2}}, & |E|<a,\\[8pt]
\dfrac{\mathrm{sgn} E}{\sqrt{E^2-a^2}}, & |E|>a.\end{cases}
\label{eq:1d:lip}
\end{equation}
The two cases are fixed by two different limiting arguments, which meet at
$E=\mp a$: for $|E|<a$ the sign of $\pm i(a^2-E^2)^{-1/2}$ is transported from
the negative imaginary axis through the lower half-plane, where $G_a$ is
analytic; for $|E|>a$ the value is real and its sign is transported along the
real axis from $\pm\infty$ on the respective side, the two half-lines
$E<-a$ and $E>a$ being connected only through the cut. The boundary value
therefore changes from real to imaginary at $E=-a$ (and back at $E=a$) with no
contradiction between the two arguments.
In particular
\begin{equation}
\boxed{\;\mathrm{Re}\, G_a(E-i0)\equiv0\quad\text{for }|E|<a\;}
\label{eq:1d:ReG}
\end{equation}
identically on the whole band, not merely at one point. Equivalently: the finite
Hilbert transform of the arcsine law vanishes on its own support, the classical
statement that the arcsine measure is the equilibrium measure of the interval. Equation~\eqref{eq:1d:ReG} is the single input that, averaged, produces
the open-channel window in every higher dimension
(Sec.~\ref{sec:2d}).

\paragraph{Every root.} A root needs $G_a(z)=i/g$; squaring,
$1/(z^2-a^2)=-1/g^2$, so $z^2=a^2-g^2\in\mathbb R$ and every root lies on the
real or the imaginary axis ($z=\pm a$ is excluded, $|G_a|\to\infty$ there).
Checking the three candidate sets:
\begin{itemize}
\item \emph{Band, lower lip} ($|E|<a$): $F=1-g(a^2-E^2)^{-1/2}=0$ gives
\begin{equation}
\boxed{\;E^2=a^2-g^2\;}
\label{eq:1d:traj}
\end{equation}
\item \emph{Outside the band}: $\mathrm{Im}\, F=g\,\mathrm{sgn} E\,(E^2-a^2)^{-1/2}\neq0$, no root.
\item \emph{Negative imaginary axis}: $1-g(y^2+a^2)^{-1/2}=0$ gives $y^2=g^2-a^2$,
which requires $g>a$. On the positive imaginary axis $F=1+g(y^2+a^2)^{-1/2}>0$.
\end{itemize}
The roots are collected in Table~\ref{tab:1d:roots}.
\begin{table}[h]
\caption{Roots of the one-dimensional secular equation $1=-igG(E)$ and the
state of the $\Sigma$ ($\mathcal{PT}$) symmetry, with $a=4t|\cos(K/2)|$. The two
real roots below threshold are CPA zeros on the lower lip of the cut, not
eigenvalues.}
\label{tab:1d:roots}
\begin{center}
\begin{tabular}{lll}
\toprule
coupling & roots & $\Sigma$ \\
\midrule
$g<a$ & two real, $E=\pm\sqrt{a^2-g^2}$, on the lower lip & broken \\
$g=a$ & one double root at $E=0$ & the transition \\
$g>a$ & one, $E=-i\sqrt{g^2-a^2}$: bound, decaying & restored \\
\bottomrule
\end{tabular}
\end{center}
\end{table}
The threshold is
\begin{equation}
\boxed{\;g_c=a=4t|\cos(K/2)|=\frac1{\pi\rho_a(0)}\;}
\label{eq:1d:gc}
\end{equation}
the last equality using $\rho_a(0)=1/(\pi a)$. It is the form $g_c=1/\pi\rho_2(0)$
that survives in higher dimensions.

\paragraph{The transition is an exceptional point.}
On the disc $|z|<a$ the lip function continues analytically as
$F(z)=1-g\,(a^2-z^2)^{-1/2}$ (principal root), an even analytic function. At
$g=g_c$,
\begin{equation}
F(0)=0,\qquad F'(0)=0,\qquad F''(0)=-\frac{g}{a^3}\neq0 ,
\end{equation}
so the root at $E=0$ is exactly double: a second-order exceptional point of the
continued secular function, at which the two CPA zeros merge. Solving
Eq.~\eqref{eq:1d:traj} near threshold,
\begin{equation}
a^2-g^2=(a-g)(a+g)\simeq2a(g_c-g)
\quad\Longrightarrow\quad
E(g)\simeq\pm\sqrt{2a}\,\sqrt{g_c-g} ,
\label{eq:1d:sqrt}
\end{equation}
the square-root branch point in the coupling. The analytic continuation of the
unnormalized $c$-norm of the complex-scaled state is proportional to
$-G'_{\rm cont}(E)$ and vanishes at the collision, since $G'_{\rm cont}(0)=0$;
it is not the norm of an eigenvector of the infinite-volume Hamiltonian.

\paragraph{What the sub-threshold roots are.}
For a rank-one interaction $U|\beta\rangle\langle\beta|$ the Born series sums
to $T(z)=\{U/[1-U\,G(z)]\}\,|\beta\rangle\langle\beta|$, and the $S$-matrix is
the identity in every channel orthogonal to $|\beta\rangle$. In the contact
channel the on-shell $S$-matrix eigenvalue is
$s(E)=1-2\pi i\,\rho_2(E)\,T(E+i0)$, the standard $S=1-2\pi iT$ with the channel
density of states $\rho_2(E)=L^{-1}\sum_q\delta(E-\varepsilon_2(q))$ made
explicit; since $G(E+i0)-G(E-i0)=-2\pi i\rho_2(E)$,
\begin{equation}
s(E)=\frac{1-U\,G(E+i0)-2\pi i\rho_2(E)\,U}{1-U\,G(E+i0)}
=\frac{1-U\,G(E-i0)}{1-U\,G(E+i0)}=\frac{F(E-i0)}{F(E+i0)},
\label{eq:1d:sratio}
\end{equation}
with $F(z)=1-U\,G(z)$ the secular function of Eq.~\eqref{eq:1d:KS}. Inside the band, using
Eq.~\eqref{eq:1d:lip} and $U=-ig$,
\begin{equation}
s(E)=\frac{v-g}{v+g},\qquad v\equiv\sqrt{a^2-E^2}=\text{relative group velocity},
\label{eq:1d:smatrix}
\end{equation}
so the real roots of Eq.~\eqref{eq:1d:traj} are exactly the zeros of $s$: an
incoming pair at that energy is completely absorbed (coherent perfect
absorption), the time reverse of a lasing threshold. At $g=g_c$ the zero is
double and the absorption line flattens, $|s|^2\propto E^4$. These roots are not
eigenvalues; the bound state above threshold is.

\paragraph{$K\to\pi$: the $\eta$-pairing point.} $a(K)\to0$: the pair band is
flat, $G=1/z$, and the only root is $E=-ig$ for every $g>0$. So $g_c=0$ and there
is no coalescence: an exceptional point requires $a>0$. This sector is excluded
throughout.

\paragraph{$K=0$ is the extremal sector.} $g_c(K)=4t|\cos(K/2)|$ is maximal at
$K=0$, so the $K=0$ sector is the last to bind as $g$ grows, and $4t$ is the
largest threshold in the problem. Sectors near $K=\pi$ bind at arbitrarily small
$g$, which is why the many-pair threshold collapses.

\paragraph{What one dimension does not settle.}
In one dimension $\mathrm{Re}\, G$ vanishes on the \emph{entire} band, so the condition
that permits real roots is free everywhere, and a pair of them exists for every
$g<g_c$ down to $g\to0$. Both features are accidents of $d=1$. In two dimensions
the same set is only a sub-interval of the band, and it can be empty; deciding
when it is non-empty is the content of the rule proved in the higher-dimensional
parts. This is why the 1D result, although published and reproduced here twice,
does not by itself establish anything about the transition in $d\ge2$.
\endgroup
\begingroup\let\section\subsection \let\subsection\subsubsection
\section{Two dimensions}
\label{sec:2d}

Nothing in the resolvent route of Sec.~\ref{sec:1d} used the dimension: the
rank-one reduction, the secular equation, the lower-lip prescription and the
criterion $g_c=1/\pi\rho_2(0)$ are all dimension-blind. Only the pair density of
states changes. Here we compute it in two dimensions, find the condition under
which the threshold is an exceptional point, and give the threshold in closed
form.

\paragraph{The pair band splits into two cosines.}
At $\bm K=0$ on the anisotropic square lattice, $\varepsilon(\bm k)=-2t_x\cos k_x-2t_y\cos k_y$
and the partner momentum is $-\bm k$, so the pair band is twice the
single-particle band:
\begin{equation}
\varepsilon_2(\bm k)=2\varepsilon(\bm k)=-4t_x\cos k_x-4t_y\cos k_y\equiv X+W .
\label{eq:2d:band}
\end{equation}
At general $\bm K$ the same substitution as in 1D, applied direction by
direction, replaces $t_i$ by $\tau_i=t_i|\cos(K_i/2)|$; we write $t_i$ for
$\tau_i$ below and take $t_x\ge t_y>0$. Define
\begin{equation}
X=-4t_x\cos k_x \ \ (\text{half-width } a=4t_x),\qquad
W=-4t_y\cos k_y \ \ (|W|\le b=4t_y).
\end{equation}
$X$ and $W$ depend on different momenta, hence are \emph{independent} random
variables, and each is an arcsine variable of the corresponding half-width. Two
immediate consequences:
\begin{equation}
\rho_2=\rho_a*\rho_b
\qquad\text{and}\qquad
\boxed{\;G(z)=\big\langle G_a(z-W)\big\rangle_W\;}
\label{eq:2d:slice}
\end{equation}
the second following by conditioning on $W$ and recognising the inner
average over $k_x$ as the one-dimensional Green function
Eq.~\eqref{eq:1d:G} evaluated at the shifted energy $z-W$. Explicitly,
\begin{equation}
G(z)=\int_{-b}^{b}\frac{dW}{\pi\sqrt{b^2-W^2}}\;\frac1{(z-W-a)^{1/2}(z-W+a)^{1/2}} .
\label{eq:2d:explicit}
\end{equation}
The two factors take principal roots separately, as in the one-dimensional
definition. Fubini applies for $\mathrm{Im}\, z\neq0$, where the integrand is bounded by
$1/|\mathrm{Im}\, z|$.

Equation~\eqref{eq:2d:slice}
holds in any dimension with $W=-4\sum_{i\ge2}t_i\cos k_i$; the function $G_a$ is
always the same. Only the \emph{law of $W$} changes: a point mass at $0$ in 1D,
an arcsine law in 2D, the two-dimensional lattice density of states in 3D, and
the $(d-1)$-dimensional one in general. Its \emph{support} will fix the interval
below; its \emph{shape} will fix the numbers.

\paragraph{The open-channel window.}
By Eq.~\eqref{eq:1d:ReG}, each term of the average in Eq.~\eqref{eq:2d:slice} is
purely imaginary whenever its argument stays inside the one-dimensional band.
That happens for every $W$ in the support precisely when $|E|+b<a$. Hence
\begin{equation}
\boxed{\;\mathrm{Re}\, G(E-i0)\equiv0\quad\text{on}\quad |E|<a-b=4|t_x-t_y|\;}
\label{eq:2d:ipv}
\end{equation}
We call this set \emph{the open-channel window}, or simply the window: every
channel of the relative motion is open there. Equivalently, the finite Hilbert transform of $\rho_2$ vanishes
identically there. The limit $\eta\to0^+$ may be taken inside the average by
dominated convergence: for $\zeta=E-W-i\eta$ with $\eta\ge0$ and
$|\mathrm{Re}\,\zeta|\le|E|+b<a$,
\begin{equation}
|\zeta^2-a^2|=|\zeta-a|\,|\zeta+a|\ \ge\ a^2-(\mathrm{Re}\,\zeta)^2\ \ge\ a^2-(|E|+b)^2>0,
\label{eq:2d:bound}
\end{equation}
so $|G_a(\zeta)|\le[a^2-(|E|+b)^2]^{-1/2}$ uniformly in $\eta$ and $W$.

An interval, and not just a point, is what matters. Bipartiteness ($\varepsilon_2(\bm k+\bm\pi)=-\varepsilon_2(\bm k)$, hence $\rho_2$
even) already gives $\mathrm{Re}\, G(0)=0$ at the single energy $E=0$, which fixes a real
threshold. It does not permit a merger. The secular equation $1=UG(E-i0)$ is
complex, i.e.\ two real conditions on one real unknown:
\begin{equation}
\underbrace{\mathrm{Re}\, G(E-i0)=0}_{\text{where a real root may sit}}
\qquad\text{and}\qquad
\underbrace{g\,\pi\rho_2(E)=1}_{\text{at which coupling}} .
\label{eq:2d:two}
\end{equation}
In the window the first is satisfied identically, so only the second remains:
one real equation, one real unknown, solvable on a continuum of energies. Outside
the window the first fails and no real root exists.

\paragraph{The two real roots and their collision.}
In the window,
\begin{equation}
\pi\rho_2(E)=\Big\langle\big[a^2-(E-W)^2\big]^{-1/2}\Big\rangle_W .
\label{eq:2d:rho}
\end{equation}
The integrand $x\mapsto(a^2-x^2)^{-1/2}$ is even and convex on $(-a,a)$, and the
law of $W$ is symmetric; a convolution of a convex function with a probability
measure is convex, so $\pi\rho_2$ restricted to the window is even and
convex, with its minimum at $E=0$ and increasing in $|E|$.

Consequently $g\pi\rho_2(E)=1$ has a symmetric pair of solutions $\pm E_0(g)$
which move inward as $g$ grows, and they meet at the centre exactly when
$g\pi\rho_2(0)=1$:
\begin{equation}
\boxed{\;g_c=\frac1{\pi\rho_2(0)}\;}
\label{eq:2d:criterion}
\end{equation}
Read energy by energy, $g(E)=1/[\pi\rho_2(E)]$ is the coupling at which an
absorption zero sits at $E$; it is maximal at $E=0$, so the band centre is the
last energy able to support one, and $g_c=\max_Eg(E)$.

The collision is an exceptional point of the continued secular function, at
which the two CPA zeros merge. On the disc $|z|<a-b$ define
\begin{equation}
\Pi(z)=\Big\langle\big[a^2-(z-W)^2\big]^{-1/2}\Big\rangle_W ,
\end{equation}
analytic there and even in $z$ (evenness by the symmetry of $W$). It agrees with
$-iG$ on the lower half-disc and on the lip, so $F(z)=1-g\Pi(z)$ is the analytic
continuation of the lip function. At $g=g_c$,
\begin{equation}
F(0)=0,\qquad F'(0)=0,\qquad F''(0)=-g\,\Pi''(0)\neq0 ,
\end{equation}
so the root is exactly double; the analytic continuation of the $c$-norm of the
complex-scaled state, proportional to $-G'_{\rm cont}(0)=-i\Pi'(0)$, vanishes; and
\begin{equation}
E_0(g)^2=\frac{2(g_c-g)}{g\,g_c\,\Pi''(0)}+O\big((g_c-g)^2\big)
\quad\Longrightarrow\quad |E_0|\propto\sqrt{g_c-g} ,
\label{eq:2d:sqrt}
\end{equation}
with the partner branch $E=-iy$, $y\propto\sqrt{g-g_c}$, above threshold. The
exponent $1/2$ on both sides is the measurable signature.

\paragraph{The threshold in closed form.}
Set $E=0$ in Eq.~\eqref{eq:2d:rho} and use the arcsine law of $W$:
\begin{equation}
\pi\rho_2(0)=\int_{-b}^{b}\frac{dW}{\pi\sqrt{b^2-W^2}}\,\frac1{\sqrt{a^2-W^2}} .
\end{equation}
Substituting $W=b\sin\theta$, the arcsine weight cancels the Jacobian exactly:
\begin{equation}
\pi\rho_2(0)=\frac1\pi\int_{-\pi/2}^{\pi/2}\frac{d\theta}{\sqrt{a^2-b^2\sin^2\theta}}
=\frac{2}{\pi a}\int_0^{\pi/2}\frac{d\theta}{\sqrt{1-\kappa^2\sin^2\theta}}
=\frac{2\,\ellK(\kappa)}{\pi a},\qquad \kappa=\frac ba=\frac{t_y}{t_x},
\end{equation}
with $\ellK$ the complete elliptic integral of the first kind in the modulus
convention. With $a=4t_x$,
\begin{equation}
\boxed{\;\pi\rho_2(0)=\frac{\ellK(t_y/t_x)}{2\pi t_x},
\qquad
g_c=\frac{2\pi t_x}{\ellK(t_y/t_x)}\;}
\label{eq:2d:gc}
\end{equation}
For $\kappa\to0$, $\ellK\to\pi/2$ and $g_c\to4t_x$, the one-dimensional
Eq.~\eqref{eq:1d:gc}; for $\kappa\to1$, $\ellK$ diverges logarithmically and
$g_c\to0$. The elliptic integral appears \emph{because} $W$ is arcsine-distributed; a
different transverse law would give a different special function, which is what
happens in three dimensions.

The same integral is a phasor average. Equation~\eqref{eq:2d:gc} can
also be read as a statement about the planar phasor $S=t_xe^{ik_x}+t_ye^{ik_y}$
of Sec.~\ref{sec:am}, whose modulus is $|S|=t_x|1+\kappa e^{i\theta}|$ with
$\theta=k_y-k_x$ uniform on $(0,2\pi)$. In that language the elliptic integral
is the \emph{two-step formula}
\begin{equation}
\Big\langle\big|1+re^{i\theta}\big|^{-1}\Big\rangle_\theta=\frac2\pi\,\ellK(r)
\quad(0\le r<1),
\qquad
\Big\langle\big|1+re^{i\theta}\big|^{-1}\Big\rangle_\theta=\frac{2}{\pi r}\,\ellK(1/r)
\quad(r>1).
\label{eq:2d:twostep}
\end{equation}
To see this, expand $|1+re^{i\theta}|^{-1}=(1+re^{i\theta})^{-1/2}(1+re^{-i\theta})^{-1/2}$
by the binomial series; the average over $\theta$ keeps only the diagonal
terms, $\sum_n[(1/2)_n/n!]^2r^{2n}={}_2F_1(\tfrac12,\tfrac12;1;r^2)=\tfrac2\pi\ellK(r)$
\cite[Eq.~19.5.1]{S_DLMF}, and the case $r>1$ follows from
$|1+re^{i\theta}|=r\,|1+r^{-1}e^{-i\theta}|$. Since
$\pi\rho_2(0)=\langle(4|S|)^{-1}\rangle$ [Eq.~\eqref{prop:am:gc}],
Eq.~\eqref{eq:2d:twostep} at $r=\kappa$ reproduces Eq.~\eqref{eq:2d:gc}; it is
used again for the isotropic cubic lattice in Sec.~\ref{sec:3d}.

\paragraph{Isotropy.}
At $t_x=t_y$ the interval Eq.~\eqref{eq:2d:ipv} shrinks to the single point
$E=0$, and two things happen together that are the same thing. First,
$\mathrm{Re}\, G(E-i0)\neq0$ for every $E\neq0$ (the vanishing on $|E|<a-b$ is
also necessary: Eq.~\eqref{prop:am:ipv} of Sec.~\ref{sec:am} gives
$\mathrm{Re}\,G(E-i0)>0$ for $E>a-b$), so there is no pair of real roots and
nothing can merge; the threshold degenerates into a single root crossing $E=0$,
with no exceptional point. Second, the saddle van~Hove singularity of the square
lattice, which sits at $\pm4|t_x-t_y|$, reaches the band centre, $\rho_2(0)$
diverges logarithmically, and $g_c=1/\pi\rho_2(0)=0$: the pair is bound at any
loss rate. This is not a bandwidth effect: from the chain to the isotropic
square lattice the pair bandwidth doubles, from $8t$ to $16t$, and yet
$\rho_2(0)$ diverges. Anisotropy is therefore the control parameter, and
Eq.~\eqref{eq:2d:gc} interpolates: every anisotropic square lattice has an
exceptional point at a finite $g_c\le4t_x$, and the isotropic one has none.

\paragraph{Non-zero total momentum.}
All of the above used $\tau_i=t_i|\cos(K_i/2)|$, so the criterion applies sector
by sector, and the isotropic square lattice, which has no exceptional point at
$\bm K=0$, has one at other momenta. At $\bm K=(0,\pi/2)$ the effective hoppings
are $\tau=(t,t/\sqrt2)$, so $\kappa=1/\sqrt2$ and
\begin{equation}
g_c=\frac{2\pi t}{\ellK(1/\sqrt2)}=\frac{8\pi^{3/2}t}{\Gamma(1/4)^2}\simeq\gcFiniteKTwoD\,t ,
\end{equation}
using $\ellK(1/\sqrt2)=\Gamma(1/4)^2/(4\sqrt\pi)$. Any statement of the form
``the isotropic lattice has no transition'' must therefore carry the qualifier
\emph{at zero total momentum}.

The higher-dimensional parts inherit Eqs.~\eqref{eq:1d:ReG},
\eqref{eq:2d:slice} and \eqref{eq:2d:criterion} unchanged, together with the
fact that a double root requires $\mathrm{Re}\,G\equiv0$ on an interval rather
than at a point. Two things are left open: in $d\ge3$ the support of $W$ is a
sum of several arcsine variables, so whether the window is non-empty
becomes a triangle-type inequality rather than $t_x\neq t_y$; and the real
roots, which in $d\le2$ exist for every $g<g_c$, need not do so in higher
dimensions.
\endgroup
\begingroup\let\section\subsection \let\subsection\subsubsection

\section{Three dimensions}
\label{sec:3d}

The rank-one reduction, the lower-lip prescription and the criterion of
Secs.~\ref{sec:1d} and~\ref{sec:2d} carry over unchanged: eigenvalues off the
free spectrum solve
\begin{equation}
1=U\,G(E),\qquad G(z)=\int\frac{\rho_2(\varepsilon)\,d\varepsilon}{z-\varepsilon},
\qquad U=-ig,
\label{eq:3d:secular}
\end{equation}
roots on the band are read on the lower lip $G(E-i0)$, where they are zeros of
the pair $S$-matrix [Eq.~\eqref{eq:1d:smatrix}], and at total momentum $\bm K$
every hopping is replaced by $\tau_i=t_i|\cos(K_i/2)|$. Only the pair density of
states is new. On the simple cubic lattice at $\bm K=0$ the pair band is
\begin{equation}
\varepsilon_2(\bm k)=-4t_x\cos k_x-4t_y\cos k_y-4t_z\cos k_z ,
\qquad |\varepsilon_2|\le\Lambda=4(t_x+t_y+t_z).
\label{eq:3d:pairband}
\end{equation}
We order the hoppings $t_1\ge t_2\ge t_3>0$ and split off the largest,
\begin{equation}
\varepsilon_2=X+W,\qquad X=-4t_1\cos k_1\ \ (a\equiv4t_1),\qquad
W=-4t_2\cos k_2-4t_3\cos k_3,\ \ |W|\le w\equiv4(t_2+t_3).
\label{eq:3d:split}
\end{equation}
$X$ and $W$ are independent, so the slice representation Eq.~\eqref{eq:2d:slice}
applies,
\begin{align}
G(z)&=\big\langle G_a(z-W)\big\rangle_W,
\label{eq:3d:slice}\\
G_a(z)&=\frac{1}{(z-a)^{1/2}(z+a)^{1/2}},\qquad G_a(-iy)=\frac{i}{\sqrt{y^2+a^2}},
\label{eq:3d:G1dz}
\end{align}
with principal roots. The one structural novelty is that $W$ is a sum of
\emph{two} cosines: its law is the density of states of a two-dimensional
lattice, not an arcsine law.

\paragraph{The pair band and the open-channel window.}
On the lower lip the one-dimensional function is
\begin{equation}
G_a(E-i0)=
\begin{cases}
i\,(a^{2}-E^{2})^{-1/2}, & |E|<a,\\[2pt]
\mathrm{sgn}(E)\,(E^{2}-a^{2})^{-1/2}, & |E|>a,
\end{cases}
\label{eq:3d:G1d}
\end{equation}
purely imaginary inside its band [Eq.~\eqref{eq:1d:ReG}]. For $|E|<a-w$ every
slice has $|E-W|\le|E|+w<a$ and is evaluated inside its own band, so
\begin{equation}
\boxed{\ \mathrm{Re}\,G(E-i0)\equiv0,\qquad
\pi\rho_2(E)=\Big\langle \big(a^{2}-(E-W)^{2}\big)^{-1/2}\Big\rangle_{W}
\quad\text{on}\quad |E|<4\,(t_1-t_2-t_3),\ }
\label{eq:3d:ipv}
\end{equation}
the limit $\eta\to0^+$ being taken inside the average by dominated convergence
with the bound Eq.~\eqref{eq:2d:bound} ($b\to w$). Hence the window is
non-empty if and only if
\begin{equation}
\boxed{\ t_1>t_2+t_3\ }
\label{eq:3d:rule}
\end{equation}
that is, if and only if $t_1,t_2,t_3$ are \emph{not} the sides of a triangle,
degenerate triangles included; necessity is Eq.~\eqref{prop:am:ipv}. In two
dimensions the corresponding condition $t_x\ne t_y$ has codimension one; three
is the first dimension in which it is an inequality with open regions of
parameter space on both sides. The edge of the window is a van~Hove
energy: the critical point $k_1=0$, $k_2=k_3=\pi$ of the pair band sits at
$-4(t_1-t_2-t_3)$, so the window closes exactly when that saddle reaches
the band centre. In three dimensions a saddle produces a \emph{kink} in the
density of states (finite value, square-root cusp), not the logarithmic
divergence of two dimensions, and everything peculiar to three dimensions below
follows from this.

\paragraph{Inside the window: the collision and its exponent.}
With $U=-ig$ the complex equation~\eqref{eq:3d:secular} splits into
$\mathrm{Re}\,G(E-i0)=0$ and $g\,\pi\rho_2(E)=1$; in the window the first
holds identically, and a real root requires only
\begin{equation}
g\,\pi\rho_2(E)=1,\qquad |E|<4(t_1-t_2-t_3).
\label{eq:3d:real}
\end{equation}
The kernel $x\mapsto(a^2-x^2)^{-1/2}$ is even and convex and the law of $W$ is
symmetric, so $\pi\rho_2$ is even and convex in the window with its
minimum at $E=0$. Equation~\eqref{eq:3d:real} therefore has a symmetric pair of
solutions $\pm E_0(g)$ that move inward as $g$ grows and meet at the centre at
\begin{equation}
g_c=\frac{1}{\pi\rho_2(0)} .
\label{eq:3d:criterion}
\end{equation}
On the disc $|z|<4(t_1-t_2-t_3)$ the function
$\Pi(z)=\langle(a^2-(z-W)^2)^{-1/2}\rangle_W$ is analytic and even, continues the
lip function as $F(z)=1-g\,\Pi(z)$, and has
\begin{equation}
\Pi''(0)=\big\langle (a^{2}+2W^{2})(a^{2}-W^{2})^{-5/2}\big\rangle_{W}>0 .
\label{eq:3d:Pipp}
\end{equation}
At $g=g_c$ we have $F(0)=F'(0)=0\neq F''(0)$: the root is exactly double, an
exceptional point of the continued secular function; the analytic continuation
of the $c$-norm of the complex-scaled state, proportional to
$-G'_{\rm cont}(0)=-i\Pi'(0)$, vanishes, and
\begin{equation}
E_0^{2}=\frac{2(g_c-g)}{g\,g_c\,\Pi''(0)}+O\big((g_c-g)^{2}\big),
\qquad
y^{2}=\frac{2(g-g_c)}{g\,g_c\,\Pi''(0)}+O\big((g-g_c)^{2}\big),
\label{eq:3d:sqrt}
\end{equation}
with $E=-iy$ above threshold: the square-root law on both sides.

\paragraph{Closed form for $(t_1,t_\perp,t_\perp)$.}
Write $\lambda=t_\perp/t_1$; Eq.~\eqref{eq:3d:rule} reads $\lambda<1/2$. For
$\lambda<1/2$, Eq.~\eqref{eq:3d:ipv} at $E=0$ gives
$\pi\rho_2(0)=\langle(16t_1^2-W^2)^{-1/2}\rangle$. The same formula holds at the
boundary $\lambda=1/2$, where the window is empty and $\pi\rho_2(0)$ is
the vertical limit $\lim_{y\to0^+}[-iG(-iy)]$ of Eq.~\eqref{prop:am:gc}: on the
imaginary axis $|G_a(-W-i\eta)|\le(a^2-W^2)^{-1/2}$, which behaves as
$|\bm k|^{-1}$ near the two points where $|W|=a$ and is integrable over
$(k_2,k_3)$, so dominated convergence applies. Since $|W|\le4t_1$, the binomial
series of $(16t_1^2-W^2)^{-1/2}$ in powers of $(W/4t_1)^2$ has non-negative
terms and may be averaged term by term. The transverse moments are
\begin{equation}
\big\langle(\cos k_2+\cos k_3)^{2n}\big\rangle=\frac{1}{4^{n}}\binom{2n}{n}^{2},
\label{eq:3d:moment}
\end{equation}
because only even powers of each cosine survive,
$\langle\cos^{2m}k\rangle=4^{-m}\binom{2m}{m}$, and
$\binom{2n}{2j}\binom{2j}{j}\binom{2n-2j}{n-j}=\binom{2n}{n}\binom{n}{j}^2$ sums
over $j$ to $\binom{2n}{n}^2$ by Vandermonde. With
$4^{-n}\binom{2n}{n}=(1/2)_n/n!$ this gives
$\langle(W/4t_1)^{2n}\rangle=(4\lambda^2)^n[(1/2)_n/n!]^2$, and Clausen's
identity \cite[Eq.~16.12.2]{S_DLMF} followed by the quadratic transformation
\cite[Eq.~15.8.18]{S_DLMF}, both at $a=b=1/4$, give
\begin{align}
\pi\rho_2(0)&=\frac{1}{4t_1}\sum_{n\ge0}\Big[\frac{(1/2)_n}{n!}\Big]^{3}(4\lambda^{2})^{n}
=\frac{1}{4t_1}\,{}_3F_2\big(\tfrac12,\tfrac12,\tfrac12;1,1;4\lambda^{2}\big)
\label{eq:3d:3F2}\\
&=\frac{1}{4t_1}\Big[{}_2F_1\big(\tfrac14,\tfrac14;1;4\lambda^2\big)\Big]^{2}
\label{eq:3d:clausen}\\
&=\frac{1}{4t_1}\Big[{}_2F_1\big(\tfrac12,\tfrac12;1;k^2\big)\Big]^{2}
=\frac{\ellK(k)^2}{\pi^2t_1},\qquad
k^{2}=\tfrac12\Big[1-\sqrt{1-4\lambda^{2}}\Big].
\label{eq:3d:modulus}
\end{align}
The ${}_3F_2$ is a perfect square, which is what makes a closed form possible.
The modulus solves $4\lambda^2=4k^2(1-k^2)$; the quadratic transformation holds
for $k^2\le1/2$, and since its left side is symmetric under $k^2\to1-k^2$ this
selects the root uniquely. Hence
\begin{equation}
\boxed{\ g_c=\frac{\pi^{2}t_1}{\ellK(k)^{2}},\qquad
k^{2}=\tfrac12\Big[1-\sqrt{1-4t_\perp^{2}/t_1^{2}}\Big],\qquad t_\perp\le t_1/2 .\ }
\label{eq:3d:gc}
\end{equation}
As $t_\perp\to0$, $\ellK\to\pi/2$ and $g_c\to4t_1$, the one-dimensional value
Eq.~\eqref{eq:1d:gc}. At $t_\perp=t_1/4$ the modulus is $k=\sin15^\circ$ and
$g_c=\gcWinThreeD\,t_1$, an exceptional point with exponent $1/2$. At the
boundary $t_\perp=t_1/2$, $k=1/\sqrt2$ and
$\ellK(1/\sqrt2)=\Gamma(1/4)^2/(4\sqrt\pi)$; equivalently, the series converges
at its endpoint (terms decay as $n^{-3/2}$) and Gauss's summation theorem gives
${}_3F_2(\tfrac12,\tfrac12,\tfrac12;1,1;1)=\Gamma(1/4)^4/(4\pi^3)$, so
$g_c=16\pi^3t_1/\Gamma(1/4)^4=\gcCloseThreeD\,t_1$, a threshold that is
\emph{not} exceptional.

\paragraph{Outside the window: a threshold that is not exceptional.}
When $t_1<t_2+t_3$ the hoppings form a triangle, the window is empty and
no two roots coexist on the lower lip, so nothing can coalesce there (we make no
statement about the continued sheet). A threshold survives at
Eq.~\eqref{eq:3d:criterion}, but a single root crosses $E=0$ linearly, with a
slope fixed by geometry. Let $R=t_2\bm\omega_2+t_3\bm\omega_3$ with independent
uniform planar unit vectors $\bm\omega_i$. The map
$(\theta_2,\theta_3)\mapsto R$ has Jacobian $t_2t_3|\sin(\theta_3-\theta_2)|$ and
two mirror-image preimages for $|t_2-t_3|<|R|<t_2+t_3$, so
\begin{equation}
p_R(\bm x)=\frac{1}{4\pi^{2}\,\Delta(|\bm x|,t_2,t_3)} ,
\label{eq:3d:pR}
\end{equation}
with $\Delta$ the area of the triangle of sides $|\bm x|,t_2,t_3$ (the
normalisation follows from Heron's formula with $u=|\bm x|^2$). The slope
relation Eq.~\eqref{prop:am:slope}, $\mathrm{Re}\,G'(0)=\tfrac{\pi}{8}p_S(\bm0)$
for $S=t_1\bm\omega_1+R$, with $p_S(\bm0)=p_R(t_1)$, then gives
\begin{equation}
\boxed{\ \mathrm{Re}\,G'(0)=\frac{1}{32\pi\,\Delta(t_1,t_2,t_3)}\ },
\qquad
16\Delta^{2}=(t_1{+}t_2{+}t_3)(-t_1{+}t_2{+}t_3)(t_1{-}t_2{+}t_3)(t_1{+}t_2{-}t_3).
\label{eq:3d:slope}
\end{equation}
Its hypothesis, $p_S$ continuous and positive at the origin, holds because for
$t_1<t_2+t_3$ and small $|\bm x|$ the radii sampled by
$p_S(\bm x)=\langle p_R(\bm x-t_1\bm\omega_1)\rangle$ lie inside the open annulus
where $p_R$ is continuous and bounded. The same formula follows independently
from Sonine's integral
$\int_0^\infty u\,J_0(t_1u)J_0(t_2u)J_0(t_3u)\,du=1/(2\pi\Delta)$~\cite{S_Watson1944}.
The bound pair therefore grows linearly,
\begin{equation}
y(g)\simeq\frac{g-g_c}{g_c^{2}\,\mathrm{Re}\,G'(0)} ,
\label{eq:3d:linear}
\end{equation}
and the slope $\mathrm{Re}\,G'(0)$ diverges as the triangle degenerates at
$t_1\to t_2+t_3$.

For the isotropic cubic lattice the triangle is equilateral and there is no
exceptional point. Here $\pi\rho_2(0)=\langle(4|S|)^{-1}\rangle=W_3(-1)/4t$, with
$W_3(-1)=\langle|S|^{-1}\rangle$ for three unit steps. Conditioning on
$R=\bm\omega_2+\bm\omega_3$, of length $2\cos\psi$ with $\psi$ uniform on
$(0,\pi/2)$, and using the two-step formula Eq.~\eqref{eq:2d:twostep},
\begin{equation}
W_3(-1)=\frac{4}{\pi^{2}}\left[\int_{\pi/3}^{\pi/2}\ellK(2\cos\psi)\,d\psi
+\int_{0}^{\pi/3}\frac{\ellK\big(1/(2\cos\psi)\big)}{2\cos\psi}\,d\psi\right],
\label{eq:3d:W3}
\end{equation}
split where $|R|=1$. The classical closed form
$W_3(-1)=\int_0^\infty J_0(x)^3dx=3\Gamma(1/3)^6/(2^{11/3}\pi^4)$
\cite[Example~5]{S_Borwein2011} then gives
\begin{equation}
\boxed{\ g_c=\frac{4t}{W_3(-1)}=\frac{2^{17/3}\pi^{4}}{3\,\Gamma(1/3)^{6}}\,t=\gcIsoThreeD\,t\ }
\label{eq:3d:iso}
\end{equation}
with $\mathrm{Re}\,G'(0)=1/(8\sqrt3\,\pi t^2)$ from Eq.~\eqref{eq:3d:slope} at
$\Delta=\sqrt3t^2/4$.

\paragraph{What is special about three dimensions.}
Three features have no analogue in one or two dimensions, and all three follow
from the edge of the window being a kink rather than a divergence. First,
the threshold stays finite when the window closes. In two dimensions
closing the interval brings the logarithmic van~Hove singularity to the band
centre and sends $g_c$ to zero; in three dimensions $\rho_2(0)$ stays finite, and
along the family $(t_1,t_\perp,t_\perp)$ the threshold falls from $4t_1$ to
$\gcCloseThreeD\,t_1$ at closure and rises again to $\gcIsoThreeD\,t$ at
isotropy. Losing the exceptional point does not cost the threshold.

Second, the real absorption zeros exist only above a positive coupling. Let
$r_{\min}=t_1-t_2-t_3>0$, so that $4r_{\min}=a-w$, and
\begin{equation}
\Pi_{\rm edge}=\Big\langle\big(a^{2}-(4r_{\min}-W)^{2}\big)^{-1/2}\Big\rangle_{W}
=\Big\langle\big[(w+W)(2a-w-W)\big]^{-1/2}\Big\rangle_{W}.
\label{eq:3d:Piedge}
\end{equation}
The second factor is at least $2(a-w)>0$, so the integrand is singular only at
$k_2=k_3=0$, where $w+W=2t_2k_2^2+2t_3k_3^2+O(|\bm k|^4)$ has a non-degenerate
minimum; the integrand is $O(|\bm k|^{-1})$ there, integrable over two
transverse angles, and $\Pi_{\rm edge}<\infty$. For fixed $W$ the integrand of
Eq.~\eqref{eq:3d:ipv} is convex in $E$ on $[0,4r_{\min}]$ and hence bounded by
its endpoint values, so by dominated convergence
$\pi\rho_2(E)\to\Pi_{\rm edge}$ as $|E|\uparrow4r_{\min}$. Since $\pi\rho_2$ is
even and strictly convex in the window, its range there is
$[\pi\rho_2(0),\Pi_{\rm edge})$, and a pair of distinct real roots strictly inside
the window exists
exactly for
\begin{equation}
g_{\rm edge}<g<g_c,\qquad g_{\rm edge}=1/\Pi_{\rm edge}>0 .
\label{eq:3d:gedge}
\end{equation}
At $g=g_{\rm edge}$ the boundary limit used above gives
$G(\pm4r_{\min}-i0)=i\Pi_{\rm edge}$ and hence $F=0$: the two zeros sit at the
van~Hove energies. They are CPA zeros, not bound-state eigenvalues. Below
$g_{\rm edge}$ the two real roots have left through the edge of the window and
the closed lower half-plane contains no root. The
argument holds in every $d\ge3$, where at least two transverse angles are
available. In one and two dimensions the singularity sits in a single angle,
$\Pi_{\rm edge}=\infty$ (the band-edge and the logarithmic van~Hove divergence),
$g_{\rm edge}=0$, and the real roots exist for every $g<g_c$; three is the first
dimension where they do not.

Third, the onset at the boundary $t_1=t_2+t_3$ is non-analytic. The triangle
degenerates, $p_S\sim|\bm x|^{-1/2}$ at the origin and the slope
Eq.~\eqref{eq:3d:slope} diverges, so the boundary is neither an exceptional
point (no window) nor a linear crossing. The harmonic-mean form of
$g_c$, Eq.~\eqref{prop:am:gc}, still converges there, which is why the first
and third features are consistent.

\paragraph{Finite total momentum.}
Replacing $t_i$ by $\tau_i=t_i|\cos(K_i/2)|$ changes the conclusions in both
directions. At $\bm K=(0,3\pi/4,3\pi/4)$ on the isotropic cubic lattice
$\tau=t\,(1,\cos\tfrac{3\pi}{8},\cos\tfrac{3\pi}{8})$ with
$2\cos(3\pi/8)=\sqrt{2-\sqrt2}<1$, so Eq.~\eqref{eq:3d:rule} holds and this
sector has an exceptional point although $\bm K=0$ has none; at
$\bm K=(0,0,\pi)$ the third hopping vanishes, the problem is the isotropic square
lattice, and $g_c=0$. A finite three-dimensional threshold is therefore a
statement at zero total momentum.
\endgroup
\begingroup\let\section\subsection \let\subsection\subsubsection

\section{The arcsine-mixture representation}
\label{sec:am}

This section derives, in arbitrary dimension $d$, one identity from which the
threshold, the condition for an exceptional point, the critical exponents and
the Bessel form of $g_c$ all follow. We use the rank-one reduction of
Sec.~\ref{sec:derive:prelim}: off the free spectrum the eigenvalues solve
\begin{equation}
1=U\,G(E),\qquad G(z)=\int\frac{\rho_2(\varepsilon)\,d\varepsilon}{z-\varepsilon},
\qquad U=-ig,\ g>0,
\label{eq:am:secular}
\end{equation}
with $F(z)=1+igG(z)$ the secular function and roots on the band read on the
lower lip $G(E-i0)$, where they are zeros of the pair $S$-matrix. The pair band
is bipartite, so $\rho_2$ is even, $G(-z^*)=-G(z)^*$ by the symmetry $\Sigma$ of
Eq.~\eqref{eq:sm:sigma}, and roots come in pairs $(z,-z^*)$. The centre value is
defined by the vertical limit
\begin{equation}
\pi\rho_2(0):=\lim_{y\to0^+}\big[-iG(-iy)\big]\in(0,+\infty],
\label{eq:am:centre}
\end{equation}
which presupposes no regularity of $\rho_2$, and $g_c=1/\pi\rho_2(0)$ is the
point where $F(0-i0)=0$ [Eq.~\eqref{eq:1d:gc}]. At total momentum $\bm K$ every
statement below holds with $t_i$ replaced by $\tau_i=t_i|\cos(K_i/2)|$; we take
$\bm K=0$ and order the hoppings $t_1\ge t_2\ge\dots\ge t_d>0$. The only
one-dimensional input is the cosine band of half-width $a$: $\varepsilon=-a\cos k$
with $k$ uniform is the cosine of a uniform angle, whose law is the arcsine law,
\begin{equation}
\Pr\{-\cos k\le x\}=\frac{\arccos(-x)}{\pi}
\quad\Longrightarrow\quad
\rho_a(\varepsilon)=\frac{1}{\pi\sqrt{a^2-\varepsilon^2}} ,
\label{eq:am:arcsine}
\end{equation}
with the Green function of Eqs.~\eqref{eq:1d:G} and~\eqref{eq:1d:lip},
\begin{equation}
G_a(z)=(z-a)^{-1/2}(z+a)^{-1/2},
\qquad
G_a(E-i0)=\begin{cases} i\,(a^2-E^2)^{-1/2}, & |E|<a,\\[2pt]
\sgn(E)\,(E^2-a^2)^{-1/2}, & |E|>a,\end{cases}
\label{eq:am:G1d}
\end{equation}
principal roots throughout, so that $\Rea G_a(E-i0)\equiv0$ inside the band and
$G_a(-iy)=i(y^2+a^2)^{-1/2}$. The observation of this section is that in every
dimension the pair energy is again the cosine of a uniform angle, times an
amplitude that is no longer constant.

\paragraph{Phasor sum and rotation invariance.}
At $\bm K=0$ the pair band is twice the single-particle dispersion,
\begin{equation}
\varepsilon_2(\bm k)=-4\sum_{i=1}^{d}t_i\cos k_i
=-4\,\Rea S(\bm k),
\qquad
S(\bm k)=\sum_{i=1}^{d}t_i\,e^{ik_i}\equiv\sum_{i=1}^{d}t_i\,\omega_i ,
\label{eq:am:phasor}
\end{equation}
a sum of $d$ phasors of fixed lengths $t_i$ and independent phases, each uniform
under the normalised Brillouin-zone measure. With $R=|S|$ and $\Theta=\arg S$,
$\varepsilon_2=-4R\cos\Theta$. Since each $e^{i\alpha}\omega_i$ is again uniform
on the circle, independently in $i$,
\begin{equation}
e^{i\alpha}S\overset{d}{=}S\qquad\text{for all }\alpha\in\mathbb R ,
\label{lem:am:rot}
\end{equation}
and a rotation-invariant law in the plane has a uniform angle independent of its
radius: $\Theta$ is uniform on $[0,2\pi)$ and independent of $R$. On the lattice,
the rotation is the diagonal shift $\bm k\mapsto\bm k+\alpha\bm 1$, under which
\begin{equation}
\varepsilon_2(\bm k+\alpha\bm 1)=-4\,\Rea\!\big[e^{i\alpha}S(\bm k)\big]
=-4\,|S(\bm k)|\cos\!\big(\alpha+\arg S(\bm k)\big)
\label{eq:am:orbit}
\end{equation}
is exactly a one-dimensional cosine band of half-width $4|S(\bm k)|$: the
Brillouin zone is foliated into one-dimensional bands of constant amplitude.
Conditionally on $R$, $4R\cos\Theta$ therefore has the arcsine law $\rho_{4R}$,
and averaging over $R$ gives the arcsine-mixture representation
\begin{align}
\varepsilon_2&\overset{d}{=}4R\cos\Theta,
\label{eq:am:mixture}\\
&\boxed{\ \rho_2=\av{\rho_{4R}}_R\ }
\label{thm:am:mixture}\\
\rho_2(\varepsilon)&=\av{\frac{\mathbf 1\{|\varepsilon|<4R\}}{\pi\sqrt{16R^2-\varepsilon^2}}}_{R} ,
\label{eq:am:mixture-explicit}
\end{align}
valid in any $d\ge1$ for any $t_i>0$: the pair density of states is a mixture of
arcsine laws, the mixing variable being the amplitude $4R$. At $d=1$, $R\equiv
t_1$ and Eq.~\eqref{thm:am:mixture} returns the one-dimensional arcsine law. The
$d$-dimensional average has become a one-parameter family of one-dimensional
problems; dimension enters only through the law of the scalar $R$.

Off the real axis, $|(z-\varepsilon)^{-1}|\le|\Ima z|^{-1}$, so Fubini's theorem
lets the Stieltjes transform pass through the average:
\begin{align}
G(z)&=\av{\int\frac{\rho_{4R}(\varepsilon)\,d\varepsilon}{z-\varepsilon}}_R
\label{prop:am:G}\\
&=\av{G_{4R}(z)}_R,\qquad \Ima z\neq0 .
\label{eq:am:G}
\end{align}
Every delicate point below is the boundary value $\Ima z\to0^-$, and each time
the limit is taken through the average we say why it may be.

The range of $R$ follows from the triangle inequality. $R$ is continuous on the
connected torus, its maximum is attained at $\bm k=0$, and its minimum is zero
exactly when $d$ segments of lengths $t_i$ can close into a (possibly
degenerate) polygon, which by the classical polygon inequality happens iff the
longest does not exceed the sum of the others. Hence the support of the law of
$R$ is $[r_{\min},r_{\max}]$ with
\begin{align}
r_{\max}&=\sum_{i=1}^{d}t_i,
\label{lem:am:range}\\
r_{\min}&=\max\Big(0,\ t_1-\sum_{i\ge2}t_i\Big),
\label{eq:am:rmin}
\end{align}
the bound $r_{\min}=t_1-\sum_{i\ge2}t_i$ being attained at $\omega_1=1$,
$\omega_{i\ge2}=-1$. $r_{\min}>0$ means that the polygon with sides
$t_1,\dots,t_d$ cannot close; for $d=3$ it is the failure of the triangle
inequality.

\paragraph{The open-channel window.}
For $0<E<4r_{\min}$ every slice contains $E$, and
$|(E-i\eta)^2-16R^2|\ge16r_{\min}^2-E^2>0$ bounds $G_{4R}(E-i\eta)$ uniformly;
by dominated convergence, $\Rea G(E-i0)=0$ and
$\pi\rho_2(E)=\av{(16R^2-E^2)^{-1/2}}_R$. For $E>4r_{\min}$ we use the quadrant
property of the one-dimensional Green function,
\begin{equation}
\Rea G_a(z)>0\qquad\text{for }\Rea z>0,\ \Ima z<0,\ a>0 .
\label{lem:am:quadrant}
\end{equation}
Writing $z\mp a=r_{1,2}e^{i\theta_{1,2}}$ with $\theta_{1,2}\in(-\pi,0)$,
$\Rea G_a>0$ iff $\theta_1+\theta_2>-\pi$; for $\Rea z=x>a$ both angles lie in
$(-\pi/2,0)$, and for $0<x<a$, with $z=x-iy$,
$\theta_1+\theta_2=-\pi+\arctan\frac{y}{a-x}-\arctan\frac{y}{a+x}>-\pi$. The
integrands $\Rea G_{4R}(E-i\eta)$ are therefore non-negative and tend to
$(E^2-16R^2)^{-1/2}$ on $\{4R<E\}$ and to zero on $\{4R>E\}$, and Fatou's lemma
gives
\begin{equation}
\Rea G(E-i0)=0\quad(E<4r_{\min}),
\qquad
\liminf_{\eta\downarrow0}\Rea G(E-i\eta)\ \ge\
\av{\frac{\mathbf 1\{4R<E\}}{\sqrt{E^2-16R^2}}}_R>0\quad(E>4r_{\min}),
\label{eq:am:ReG}
\end{equation}
positive because the law of $R$ charges every neighbourhood of $r_{\min}$
[Eq.~\eqref{eq:am:rmin}]. The limit exists and equality holds when $R$ has a
density bounded near $E/4$, the slices with $4R$ just above $E$ then contributing
$O(\eta^{1/2})$. With the symmetry $\Sigma$, the principal value vanishes exactly
on
\begin{align}
\Rea G(E-i0)\equiv0\quad&\Longleftrightarrow\quad |E|<4r_{\min},
\label{prop:am:ipv}\\
4r_{\min}&=4\big(t_1-\textstyle\sum_{i\ge2}t_i\big)_+ ,
\label{eq:am:ipv}
\end{align}
the open-channel window, which is non-empty iff $t_1>\sum_{i\ge2}t_i$, while outside it $\Rea G(E-i\eta)$
stays bounded away from zero as $\eta\downarrow0$. The inclusion is also what the
slice split of the dimensional sections gives, since the transverse energy obeys
$|W|\le4\sum_{i\ge2}t_i$; what the mixture adds is necessity. Its slices are all
centred at $E=0$, so outside the window the slices that have
already ended contribute principal values of one sign, with no cancellation.

\paragraph{Threshold, root census and exponents.}
On the imaginary axis Eqs.~\eqref{eq:am:G} and~\eqref{eq:am:G1d} give
$G(-iy)=i\Phi(y)$ with
\begin{equation}
\Phi(y)=\av{\frac{1}{\sqrt{y^2+16R^2}}}_R ,
\label{eq:am:Phi}
\end{equation}
real, positive and strictly decreasing to $\Phi(\infty)=0$, so that
$F(-iy)=1-g\Phi(y)$. The integrand increases monotonically to $(4R)^{-1}$ as
$y\downarrow0$, and monotone convergence with Eq.~\eqref{eq:am:centre} gives
\begin{align}
\pi\rho_2(0)&=\Phi(0^+)=\av{\frac{1}{4R}}_R ,
\label{prop:am:gc}\\
&\boxed{\ g_c=\Big[\av{(4R)^{-1}}_R\Big]^{-1}\ },
\label{eq:am:gc}
\end{align}
the harmonic mean of the slice half-widths, with $F(-iy)=0$ having exactly one
root $y>0$ for $g>g_c$ and none for $g<g_c$. The harmonic mean is dominated by
the narrowest slices: $g_c$ vanishes exactly when $\av{R^{-1}}$ diverges, as on
the isotropic square lattice at $\bm K=0$, where $R$ can vanish (the
$\eta$-pairing sector $\bm K=\bm\pi$, where every $\tau_i=0$ and the band is
flat, has $g_c=0$ for a different reason).

A root needs $G(z)=i/g$, hence $\Rea G(z)=0$. By Eqs.~\eqref{eq:am:G}
and~\eqref{lem:am:quadrant}, $\Rea G>0$ on the open fourth quadrant, hence by
$G(\bar z)=\overline{G(z)}$ and $\Sigma$ on every open quadrant; on the positive
imaginary axis $F(iy)=1+g\Phi(y)>0$; and on the real axis outside
the window and at its endpoints, Eq.~\eqref{eq:am:ReG} keeps $\Ima F=g\Rea G$ away from
zero. The roots therefore satisfy
\begin{equation}
F(z)=0\ \Longrightarrow\ z\in\{-iy:\ y>0\}\ \cup\ \{|E|<4r_{\min}\}\ \cup\
\{0\ \text{at}\ g=g_c\} .
\label{thm:am:census}
\end{equation}
An endpoint is a zero only if its finite boundary value satisfies
$F(E-i0)=0$, which happens at $g=g_{\rm edge}$ in the three-dimensional
calculation. Inside the window the condition is $g\,\pi\rho_2(E)=1$ with
$\pi\rho_2(E)=\av{(16R^2-E^2)^{-1/2}}$, an average of even, strictly convex
functions, so the real roots form a $\Sigma$-conjugate pair $\pm E_0(g)$ that
moves inward as $g$ grows and collides at $E=0$ when $g=g_c$. The point $E=0$ at
$g=g_c$ with the window empty is the lone crossing root of
Eq.~\eqref{eq:am:linear} below.

When the window is non-empty the lip function continues to the
even analytic function
\begin{equation}
\Pi(z)=\av{\big(16R^2-z^2\big)^{-1/2}}_R,
\qquad
\Pi(0)=\frac1{g_c},\qquad
\Pi''(0)=\av{(4R)^{-3}}_R\in(0,\infty)
\label{eq:am:Pi}
\end{equation}
on the disc $|z|<4r_{\min}$, the moment being finite because $R\ge r_{\min}>0$.
At $g=g_c$, $F=1-g\Pi$ has $F(0)=F'(0)=0\neq F''(0)$: the root at the origin is
exactly double, an exceptional point of the continued secular function, and
\begin{equation}
\boxed{\ E_0^2=\frac{2(g_c-g)}{g\,g_c\,\av{(4R)^{-3}}}+O\big((g_c-g)^2\big),
\qquad
y^2=\frac{2(g-g_c)}{g\,g_c\,\av{(4R)^{-3}}}+O\big((g-g_c)^2\big),\ }
\label{eq:am:sqrt}
\end{equation}
the square-root law on both sides; the analytic continuation of the $c$-norm
of the complex-scaled state, proportional to $-G'_{\rm cont}(0)$, vanishes. When
the window is empty ($r_{\min}=0$) there are no real roots besides
the crossing point, and $F'(0)=igG'(0)$. If $0<\Rea G'(0)<\infty$ the zero is
simple, and a single root crosses $E=0$ and descends the imaginary axis linearly,
\begin{equation}
y\simeq\frac{g-g_c}{g_c^{2}\,\Rea G'(0)}\,,\qquad g\downarrow g_c ,
\label{eq:am:linear}
\end{equation}
with exponent $1$ and no coalescence: a threshold, not an exceptional point. The
cases $\Rea G'(0)=0$ and $+\infty$, which occur on the boundary
$t_1=\sum_{i\ge2}t_i$ and on some resonant lattices in $d\le4$, are outside
Eq.~\eqref{eq:am:linear} and are treated in Sec.~\ref{sec:hd}.

\paragraph{The slope at the centre is a return density.}
From $G(-iy)=i\Phi(y)$, $G'(-iy)=-\Phi'(y)=\av{y\,(y^2+16R^2)^{-3/2}}_R>0$. Let
the planar vector $S$ have a density $p_S$, bounded and continuous on a disc
$|\bm x|\le r_0$; by rotation invariance $R$ has there the radial density
$2\pi r\,p_S(r)$. The part of the average with $R>r_0$ is at most
$y/(64r_0^3)\to0$, whatever the law of $R$ does at larger $r$; in the part with
$R<r_0$ we substitute $r=yu/4$ and apply dominated convergence with the majorant
$\sup p_S\cdot u(1+u^2)^{-3/2}$:
\begin{align}
\Rea G'(0)=-\Phi'(0^+)&=\lim_{y\to0^+}\frac{\pi}{8}\int_0^{4r_0/y}
\frac{u\,p_S(yu/4)}{(1+u^2)^{3/2}}\,du ,
\label{eq:am:slope}\\
&\boxed{\ \Rea G'(0)=\frac{\pi}{8}\,p_S(\bm 0)\ }.
\label{prop:am:slope}
\end{align}
$p_S(\bm0)$ is the density at the origin of the endpoint of the phasor sum, the
return density of the flight. Independence of the phasors gives, for the partial
sums $S_m=\sum_{i\le m}t_i\omega_i$,
\begin{equation}
p_{S_m}(\bm x)=\frac1{2\pi}\int_0^{2\pi}p_{S_{m-1}}\big(\bm x-t_me^{i\theta}\big)\,d\theta ,
\label{eq:am:convolution}
\end{equation}
and $p_{S_2}(\bm x)=1/(4\pi^2\Delta)$ with $\Delta$ the area of the triangle of
sides $t_1,t_2,|\bm x|$, positive on the open annulus $|t_1-t_2|<|\bm x|<t_1+t_2$.
By induction on $m$ — the circle $\bm x-t_me^{i\theta}$ meets the interior of the
support of $S_{m-1}$ on an open arc whenever $\bm x$ is interior to that of $S_m$,
by Eq.~\eqref{eq:am:rmin} applied to the first $m$ hoppings —
for $d\ge2$
\begin{equation}
p_{S}(\bm x)>0\quad\text{for}\ r_{\min}<|\bm x|<r_{\max},
\qquad
p_S(\bm 0)>0\quad\text{if}\ t_1<\sum_{i\ge2}t_i ,
\label{lem:am:positive}
\end{equation}
possibly $+\infty$, the origin being interior to the support exactly when
$t_1<\sum_{i\ge2}t_i$ strictly. For
$t_1>\sum_{i\ge2}t_i$ the support excludes the disc $|\bm x|<r_{\min}$ and
$-\Phi'(y)\le y/(64r_{\min}^3)\to0$ directly. With Eqs.~\eqref{eq:am:ipv},
\eqref{eq:am:sqrt} and~\eqref{eq:am:linear}, for $t_1\neq\sum_{i\ge2}t_i$,
\begin{equation}
\boxed{\ p_S(\bm 0)=0\iff\text{the polygon cannot close}\iff
t_1>\sum_{i\ge2}t_i\iff\text{exceptional point, exponent }\tfrac12,\ }
\label{cor:am:dichotomy}
\end{equation}
while $t_1<\sum_{i\ge2}t_i$ gives $0<p_S(\bm0)\le\infty$, and whenever
$0<p_S(\bm 0)<\infty$ the onset is linear, with exponent $1$ and slope
$8/(\pi g_c^2 p_S(\bm0))$. The exceptional point exists iff the flight cannot
return to its origin. The case $p_S(\bm0)=+\infty$, which occurs on resonant
lattices in $d=4$, has no coalescence but no finite slope
(Sec.~\ref{sec:hd}). On the boundary $t_1=\sum_{i\ge2}t_i$,
the window is empty and nothing coalesces, but $p_S(\bm0)$ may be
$0$, finite or $+\infty$ depending on $d$; the cases $d=3$ and $d\ge5$ are
resolved in Sec.~\ref{sec:hd} and $d=4$ remains open. In $d=3$,
Sonine's three-Bessel integral~\cite{S_Watson1944} gives $p_S(\bm0)=1/(4\pi^2\Delta)$
with $\Delta$ the area of the triangle of sides $t_1,t_2,t_3$, so
$\Rea G'(0)=1/(32\pi\Delta)$. None of this uses the regularity of $\rho_2$, which
enters only through $\Phi$, $\Pi$ and the limit~\eqref{eq:am:centre}; and it
describes the thermodynamic limit only. On a finite torus with even sides, where the pair
energies come in $\pm\varepsilon_2$ pairs, and without an exact pair zero mode, the analogue $\Phi_L(y)=N_s^{-1}\sum_{\bm k}y/(y^2+\varepsilon_2(\bm k)^2)$
of Eq.~\eqref{eq:am:Phi} vanishes at the origin instead of tending to $1/g_c$,
while its slope there, $N_s^{-1}\sum_{\bm k}\varepsilon_2(\bm k)^{-2}$, is
positive; since $\Phi_L$ also vanishes as $y\to\infty$, it has an interior
maximum, where $\Phi_L'=0$ and two roots on the imaginary axis merge. Every such
finite torus therefore shows a merger, whether or not the polygon closes
(Sec.~\ref{sec:sp}).

\paragraph{Bessel form and moments.}
The phasors are independent and uniform on their circles, so the characteristic
function of $S$ factorises; for $|\bm u|=s$, by rotation invariance,
\begin{equation}
\av{e^{i\bm u\cdot\bm S}}=\prod_{i=1}^{d}J_0(t_i s),
\qquad\text{hence}\qquad
\av{J_0(Rs)}_R=\prod_{i=1}^{d}J_0(t_i s) .
\label{eq:am:charfun}
\end{equation}
Inserting $\int_0^\infty e^{-ys}J_0(bs)\,ds=(y^2+b^2)^{-1/2}$ into
Eq.~\eqref{eq:am:Phi} and exchanging the integrals (Fubini, $|J_0|\le1$) gives
\begin{equation}
\boxed{\ \begin{gathered}
\Phi(y)=\int_0^\infty e^{-ys}\prod_{i=1}^{d}J_0(4t_i s)\,ds,\\[2pt]
\frac1{g_c}=\int_0^\infty\prod_{i=1}^{d}J_0(4t_i s)\,ds,
\qquad
\Rea G'(0)=\int_0^\infty s\prod_{i=1}^{d}J_0(4t_is)\,ds.
\end{gathered}\ }
\label{eq:am:watson}
\end{equation}
The first line holds for every $y>0$. The second line consists of Abelian limits
$y\to0^+$ of Laplace transforms and holds whenever the integrals converge,
absolutely or conditionally; since $\prod_iJ_0(4t_is)=O(s^{-d/2})$, the integral
for $1/g_c$ converges absolutely for $d\ge3$, conditionally at $d=2$ with
$t_1\neq t_2$, and diverges at $d=2$ isotropy, where $g_c=0$, while the integral
for $\Rea G'(0)$ converges absolutely only for $d\ge5$, at $d=3,4$ by cancellation
on non-resonant hopping sets, and can diverge otherwise
(Sec.~\ref{sec:hd}). The isotropic case is $1/g_c=I_d/(4t)$ with
$I_d=\int_0^\infty J_0(x)^d\,dx$.

The Bessel-product form is standard, and so are the objects in it: $S$ is
Pearson's planar random flight of $d$ steps of lengths $t_i$, whose amplitude law
$\Pr\{R<r\}=r\int_0^\infty J_1(rs)\prod_iJ_0(t_is)\,ds$ is Kluyver's formula, the
Hankel inversion of Eq.~\eqref{eq:am:charfun}, and the integrals at $y=0$ are
Watson integrals. What is used here is the physical reading of the mixture, which
holds whether or not any integral can be evaluated. The flight lives in the plane
of the hopping phases, not on the lattice; the lattice walk is related to it by
an exact moment identity. With all $t_i=1$, let $c_{2k}^{(d)}$ count closed
$2k$-step nearest-neighbour walks on $\mathbb Z^d$ and $W_d(2k)=\av{R^{2k}}$ be the
flight moments~\cite{S_Borwein2011}; then Eq.~\eqref{eq:am:mixture} and the
independence of $R$ and $\Theta$ give
\begin{align}
c_{2k}^{(d)}&=\av{\Big(\sum_i2\cos k_i\Big)^{2k}}
=4^k\av{R^{2k}}\av{\cos^{2k}\Theta},
\label{prop:am:moments}\\
c_{2k}^{(d)}&=\binom{2k}{k}\,W_d(2k),
\label{eq:am:moments}
\end{align}
the one-dimensional factor $\binom{2k}{k}=4^k\av{\cos^{2k}\Theta}$ being the
arcsine law's moment; $d=1,2$ give $c_{2k}=\binom{2k}{k}$ and $\binom{2k}{k}^2$.
\endgroup
\begingroup\let\section\subsection \let\subsection\subsubsection
\providecommand{\Rea}{\mathrm{Re}\,}
\providecommand{\av}[1]{\left\langle #1\right\rangle}
\providecommand{\alsolabel}[1]{\csname ltx@label\endcsname{#1}}

\section{Four dimensions and above}
\label{sec:hd}

Above three dimensions no closed form for $g_c$ is known, isotropic or not:
Clausen's product formula behind Eq.~\eqref{eq:3d:gc} and Sonine's
three-Bessel integral behind Eq.~\eqref{eq:3d:slope} do not continue. What
remains is the Bessel form~\eqref{eq:am:watson}, evaluated by quadrature, and
the arcsine mixture of Sec.~\ref{sec:am}, in particular
$g_c=[\av{(4R)^{-1}}]^{-1}$ with $R=|S|$, $S=\sum_it_ie^{ik_i}$
[Eq.~\eqref{prop:am:gc}], and $\Rea G'(0)=\tfrac\pi8p_S(\bm0)$
[Eq.~\eqref{prop:am:slope}]. The only new ingredient is the large-$u$
behaviour of $\prod_iJ_0(t_iu)$, which is where the parity of $d$ enters.

\begin{figure}[t]
\includegraphics[width=\textwidth]{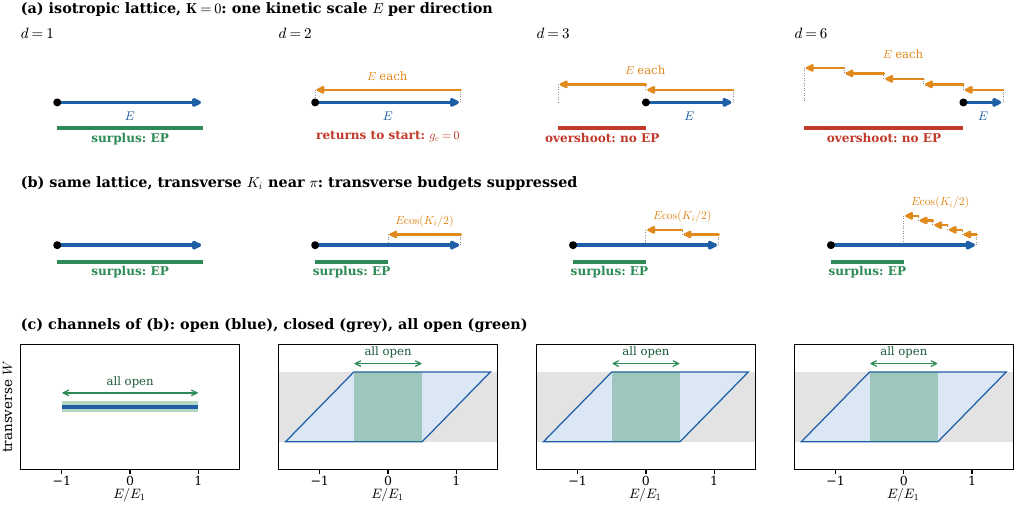}
\caption{Kinetic competition from one dimension to many. $E_i=4t_i|\cos(K_i/2)|=2J_i$ is the kinetic energy of the
pair's relative motion along direction $i$; the budgets are drawn as a chain, the dominant direction forward and
every transverse direction folded back. (a) Isotropic lattice at $\bm K=0$, one scale $E$ per direction (each
column to its own scale): in one dimension the chain stops short of its start, a kinetic surplus and an exceptional
point (EP); in two it returns exactly to the start and the pair binds at any loss; from three dimensions on it
overshoots, more with every added direction, and the threshold is not exceptional. (b) The same lattice in a
sector whose transverse momenta lie near $\pi$: the factors $\cos(K_i/2)$ suppress the transverse budgets and the
surplus, and with it the EP, returns in every dimension. (c) Channels of the sectors in (b): every channel of the
relative motion is open (green) in a window whose half-width is the surplus.}
\label{fig:dimensions}
\end{figure}

Fig.~\ref{fig:dimensions} shows the rule as the competition between kinetic
energies of the pair's relative motion that the Letter describes. With
$E_i=4t_i|\cos(K_i/2)|$ the kinetic energy along direction $i$, the transverse
budgets folded back against the dominant one stop short of the start by
$E_1-\sum_{i\ge2}E_i$, which is the half-width of the window in which every
channel is open, Eq.~\eqref{eq:am:ipv}; when they reach back past the start the
polygon of sides $E_i$ closes. On the isotropic lattice at $\bm K=0$ the
transverse budgets win from $d=2$ on, by more with every added direction
[Fig.~\ref{fig:dimensions}(a)]; total momentum near $\pi$ along the transverse
directions suppresses them and restores the surplus in every dimension
[Fig.~\ref{fig:dimensions}(b),(c)].

The rule $t_1>\sum_{i\ge2}t_i$ is dimension-blind as a formula but requires
one hopping to outweigh $d-1$ others. It is never vacuous ($t_1=1$,
$t_i=1/[2(d-1)]$ satisfies it in every $d$), but it becomes rare. Let only the
ratios of the hoppings be random, uniformly (Dirichlet) on the simplex
$\sum_it_i=1$. The rule then reads $t_{\max}>\tfrac12$, and since
$\Pr\{t_{\max}>x\}=\sum_{j\ge1}(-1)^{j-1}\binom dj(1-jx)_+^{d-1}$, in which
every term with $j\ge2$ vanishes at $x=\tfrac12$, the probability that the
lattice has a zero-momentum exceptional point is
\begin{equation}
\boxed{\ \Pr\Big\{t_{\max}>\tfrac12\Big\}=d\,2^{1-d},\ }
\label{eq:hd:fraction}\alsolabel{prop:hd:fraction}
\end{equation}
that is $1$, $3/4$, $1/2$, $5/16$, $3/16$, $7/64$, $1/16$ for $d=2,\dots,8$.
At $d=2$ the rule fails only on the line $t_1=t_2$, which is why two
dimensions look generic; from $d=4$ on it holds at most at even odds. Dimension
does not forbid the effect but demands an increasingly lopsided lattice.

This decay concerns $\bm K=0$ only. At total momentum $\bm K$ every result
holds with the effective hoppings $\tau_i=t_i|\cos(K_i/2)|$
(Sec.~\ref{sec:derive:prelim}), so for every $d$ and every set of hoppings the
pair sector has an exceptional point wherever
\begin{equation}
t_1|\cos(K_1/2)|>\sum_{i\ge2}t_i|\cos(K_i/2)| ,
\label{eq:hd:Krule}\alsolabel{prop:hd:K}
\end{equation}
an open set containing, for instance, $K_1=0$ with every $K_{i\ge2}$ close
enough to $\pi$, since $\tau_i\to0$ as $K_i\to\pi$. (The limit $K_i\to\pi$ for
all $i$ is excluded: the pair band flattens and $g_c\to0$, $\eta$-pairing.)
High dimension thus confines the exceptional point to an open set of total
momenta rather than suppressing it.

\paragraph{The isotropic lattice and the parity of $d$.}
For $t_i=t$ the Bessel form~\eqref{eq:am:watson} collapses to
\begin{equation}
\boxed{\ g_c=\frac{4t}{I_d},\qquad I_d=\int_0^\infty J_0(x)^d\,dx,\ }
\label{eq:hd:iso}
\end{equation}
so $g_c/t=4$ at $d=1$, $0$ at $d=2$ ($I_2$ diverges), $\gcIsoThreeD$ at
$d=3$ [Eq.~\eqref{eq:3d:iso}], $\gcIsoFourD$ and $\gcIsoFiveD$ at $d=4,5$, and
$\gcIsoEightD$ at $d=8$. Since $t<(d-1)t$ the polygon closes for $d\ge3$, and
by Eqs.~\eqref{cor:am:dichotomy} and~\eqref{lem:am:positive} there is no
exceptional point. The onset is linear with the finite slope of
Eq.~\eqref{eq:am:linear} wherever $p_S(\bm0)<\infty$, i.e.\ at $d=3$ and
$d\ge5$; at $d=4$ the slope diverges, but the window is still empty and
nothing coalesces.

The sequence is not monotone, $g_c$ falling from $d=3$ to $d=4$, and the
mechanism is resonance. A set of hoppings is resonant if some signed sum
vanishes, $\sum_i\sigma_it_i=0$ with $\sigma_i=\pm1$; isotropic lattices are
resonant exactly in even $d$. From $J_0(x)\sim\sqrt{2/\pi x}\,\cos(x-\pi/4)$,
\begin{equation}
\prod_{i=1}^dJ_0(t_iu)\sim\Big(\frac{2}{\pi u}\Big)^{d/2}
\frac{1}{\sqrt{\textstyle\prod_it_i}}\
\frac{1}{2^{d}}\sum_{\bm\sigma}\cos\Big(u\sum_i\sigma_it_i-\frac\pi4(n_+-n_-)\Big),
\label{eq:hd:statphase}
\end{equation}
a sum over the $2^d$ sign patterns, $n_\pm$ counting the signs of each. Every
term oscillates except those with $\sum_i\sigma_it_i=0$, which leave the
non-oscillating tail
\begin{equation}
\Big(\frac{2}{\pi u}\Big)^{d/2}\frac{2^{-d}}{\sqrt{\prod_it_i}}
\sum_{\bm\sigma:\ \sum_i\sigma_it_i=0}\cos\Big[\frac\pi4\,(n_+-n_-)\Big] .
\label{eq:hd:tail}
\end{equation}
For even $d$ the phase factor is $+1$ when
$n_+-n_-\equiv0\pmod 8$, $-1$ when $n_+-n_-\equiv4\pmod 8$, and $0$ when
$n_+-n_-\equiv2\pmod 4$, in which case the resonant pattern contributes no
leading tail; for odd $d$ it is $\pm1/\sqrt2$. On the isotropic lattice in
even $d$ every resonant pattern has $n_+=n_-=d/2$ and the coefficient is
positive.

Three consequences follow. First, $p_S(\bm0)=\frac1{2\pi}\int_0^\infty
u\prod_iJ_0(t_iu)\,du$ has an integrand $O(u^{1-d/2})$ by
Eq.~\eqref{eq:hd:statphase}, so
\begin{equation}
\boxed{\ d\ge5:\ p_S \text{ is bounded and continuous, for every set of hoppings.}\ }
\label{eq:hd:reg}
\end{equation}
At $d=4$ a resonant tail with non-zero phase factor gives $\int du/u$:
$p_S(\bm0)$, and with it $\Rea G'(0)$, diverges logarithmically, the isotropic
lattice included; at $d=3$ the divergence near the degenerate triangle is the
stronger $\int u^{-1/2}du$; non-resonant sets in $d=3,4$ converge by
cancellation. Second, on isotropic lattices $\Rea G'(0)$ therefore diverges at
$d=2$ and $d=4$ and at no higher even dimension. Third, for $d\ge6$ a positive
coefficient in Eq.~\eqref{eq:hd:tail} still \emph{enhances} $p_S$ near the
origin, the tail adding without cancellation; more weight at small $R$ means a
larger $\av{(4R)^{-1}}$ and a smaller $g_c$. That is the dip: even dimensions
sit below the odd-dimensional trend, in $g_c$, in the slope and in the sign of
the asymptotic error below. The enhancement is not a
property of resonance as such: for $d=8$ and hoppings $(5,1,1,1,1,1,1,1)$
every resonant pattern has $n_+-n_-=\mp4$, the coefficient in
Eq.~\eqref{eq:hd:tail} is negative, and the tail \emph{depletes} $p_S$ near
the origin.

\paragraph{Large $d$.}
On the isotropic lattice
\begin{equation}
\boxed{\ \frac{g_c}{4t\sqrt{d/\pi}}\ \xrightarrow[d\to\infty]{}\ 1,\qquad
8dt^2\,\Rea G'(0)\ \xrightarrow[d\to\infty]{}\ 1 .\ }
\label{eq:hd:asym}\alsolabel{prop:hd:asym}
\end{equation}
$S=t\sum_i\omega_i$ is a sum of $d$ independent, mean-zero planar unit
vectors, so by the central limit theorem $S/\sqrt d$ tends to an isotropic
Gaussian with $\av{|S|^2}=dt^2$, for which $R$ is Rayleigh-distributed with
$\av{R^{-1}}=\sqrt\pi/(t\sqrt d)$. To pass the limit through the singular
functional $R^{-1}$, the region $R<\epsilon t\sqrt d$, which contributes at
most $2\pi\epsilon t\sqrt d\,\sup p_S$, must be a fraction $O(\epsilon)$ of
$\av{R^{-1}}$ uniformly in $d$, i.e.\ $\sup p_S=O(1/dt^2)$. This holds: by
Fourier inversion of
Eq.~\eqref{eq:am:charfun}, $p_S(\bm x)\le\frac1{2\pi}\int_0^\infty
u\,|J_0(tu)|^d\,du$ for $d\ge5$, and with $0\le J_0(v)\le e^{-v^2/4}$ on
$[0,2]$ and $c=\sup_{v\ge2}|J_0(v)|<1$,
\begin{equation}
\sup_{\bm x}p_S(\bm x)\le\frac{1}{2\pi t^2}\Big[\int_0^2ve^{-dv^2/4}dv
+c^{\,d-5}\!\int_2^\infty v|J_0(v)|^5dv\Big]
=\frac{1}{\pi dt^2}\big(1+O(d\,c^{\,d})\big).
\label{eq:hd:supbound}
\end{equation}
The truncated functional $\min[R^{-1},(\epsilon t\sqrt d)^{-1}]$ then
converges by the central limit theorem, and $\epsilon\to0$ gives the first
limit. For the second, the same split of
$p_S(\bm0)$ at $v=2$ and Laplace's method on $J_0(v)^d=e^{-dv^2/4+O(dv^4)}$
give $p_S(\bm0)=(1+o(1))/(\pi dt^2)$, which Eq.~\eqref{prop:am:slope} turns
into the slope.

Only the leading orders are proved. The refinement
$g_c=4t\sqrt{d/\pi}\,[1+3/(16d)]$ is taken from the numerics, not derived; it
is accurate to better than a part in a thousand by $d=8$ and lies
\emph{below} the true $g_c$ at odd $d$ and \emph{above} it at even $d$, the
resonant enhancement being invisible to the Gaussian limit. The slopes
computed at $d=5,6,7$ bracket $1/(8dt^2)$ the same way, the even value alone
lying above. The isotropic lattice thus needs an ever stronger loss, growing
as $\sqrt d$, to bind a pair at zero momentum. The exact threshold exceeds $4t$
at the dimensions whose values are quoted above ($d=3$, $4$, $5$ and $8$) and,
by Eq.~\eqref{eq:hd:asym}, for all sufficiently large $d$; that it does so at
every $d\ge3$ is not established here. The leading term $4t\sqrt{d/\pi}$ alone falls below $4t$
for $d\le3$ and must not be used for that comparison at small $d$.
The dip between $d=3$ and $d=4$ is a small resonance-driven exception inside
this trend, not a reversal.

\paragraph{The boundary $t_1=\sum_{i\ge2}t_i$.}
Here the window is empty and Eq.~\eqref{thm:am:census} forbids real
roots other than the crossing $E=0$ at $g=g_c$: nothing coalesces and no
discrete state exists at $g_c$. The quantity to follow is
$-G'(0)$, the limit as $g\downarrow g_c$ of the $c$-norm $-G'(z_0)$ of the
bound pair at $z_0=-iy$, which vanishes at an exceptional point. There
\begin{equation}
p_S(\bm0)=\begin{cases}0, & d\ge5,\\ +\infty, & d=3,\end{cases}
\qquad \Rea G'(0)=\tfrac\pi8\,p_S(\bm0),
\label{prop:hd:bdry}
\end{equation}
three dimensions being the atypical case. For $d\ge5$, $p_S(\bm0)$ is
continuous in the hoppings [its integral converges absolutely and locally
uniformly, Eq.~\eqref{eq:hd:reg}]; it
vanishes on the open region $t_1>\sum_{i\ge2}t_i$, where $\bm0$ lies outside
the support of $S$ [Eq.~\eqref{lem:am:range}], and hence on its closure. For
$d=3$, $\Rea G'(0)=1/(32\pi\Delta)$ of Eq.~\eqref{eq:3d:slope} diverges as the
triangle degenerates, $\Delta\to0$. The case $d=4$ is marginal and left open,
the integral being only conditionally convergent. The boundary $t_1=t_2+t_3+t_4$ is resonant, but its only resonant
patterns are $\pm(+,-,-,-)$, with $n_+-n_-=\mp2$, so by
Eq.~\eqref{eq:hd:tail} the leading $u^{-2}$ tail of $\prod_iJ_0(t_iu)$
vanishes and the logarithm of the isotropic case is absent at leading order;
the next non-oscillating term, $O(u^{-3})$, gives an integrable $O(u^{-2})$ in
$u\prod_iJ_0$. Whether $p_S(\bm0)$ vanishes there is a question about a
conditionally convergent integral, and we make no claim; the convolution
identity~\eqref{eq:am:convolution}, which reads $p_S(\bm0)=p_{S_3}(t_1)$ with
$t_1$ the outer edge of the support of $S_3$, suggests a finite non-zero
value.

The case $d\ge5$ corrects a diagnostic. On the entire boundary,
\begin{equation}
-G'(0)=\lim_{g\downarrow g_c}\big[-G'(z_0)\big]=0
\quad\text{while}\quad r_{\min}=0,\qquad d\ge5 ,
\label{cor:hd:cnorm}
\end{equation}
so the limiting $c$-norm vanishes although the spectrum has no degeneracy. A
vanishing $c$-norm is necessary but not sufficient, and a certification of
an exceptional point resting on it alone is invalid in $d\ge5$; in $d\le3$ the
two agree.

Finally, the edge of the window is still the van~Hove energy of the
critical point $k_1=0$, $k_{i\ge2}=\pi$, but in $d\ge4$ the density of states
there is finite and increasingly smooth, so the singularity structure says
nothing about the threshold. Finite lattices mislead more than in three
dimensions: accidental near-zero pair levels are denser, every finite torus with
even sides and without exact zero modes shows a merger, and no
exceptional point in $d\ge4$ should be certified from a single lattice size.
\endgroup
\begingroup\let\section\subsection \let\subsection\subsubsection

\section{From finite lattices to the continuum: the one-pair spectrum}
\label{sec:sp}

The Letter follows the most-decaying levels of a fixed-momentum pair sector,
the states of most negative $\mathrm{Im}\,E$. On a finite lattice that
state has many competitors: every level with some weight on double occupancy
decays. This section shows that the competition disappears in the
thermodynamic limit. On an $L_1\times\dots\times L_d$ torus, at fixed total
momentum $\bm K$, all decay rates except one tend to zero, and the exception is
the bound pair $E=-iy_*$ of the infinite-lattice secular equation, present
exactly when $g>g_c$. Precisely: for $g\neq g_c$ and every $\eta>0$ with
$\eta\neq y_*$ there is an $L_0$ such that, whenever $\min_iL_i\ge L_0$,
\begin{equation}
\#\{n:\ \mathrm{Im}\,E_n<-\eta\}=
\begin{cases}
1, & g>g_c\ \text{and}\ \eta<y_*,\\
0, & \text{otherwise},
\end{cases}
\qquad
-\eta\le\mathrm{Im}\,E_n\le0\ \text{for every other}\ n,
\label{eq:sp:main}
\end{equation}
eigenvalues counted with algebraic multiplicity. The isolated eigenvalue
$E_b$ converges to $-iy_*$ at the rate $\sum_i\tau_i/L_i$, is purely imaginary
whenever every $L_i$ is even, and is therefore the state of most negative
$\mathrm{Im}\,E$ in its sector. Underneath the statement is an exact sum rule:
the decay rates of the sector always add up to $g$.

\paragraph{Setting.}
Take one up and one down fermion on the torus, $N_s=\prod_iL_i$ sites, with
$H$ of Eq.~\eqref{eq:sm:ray} and a total
momentum $\bm K$ on the grid $\{2\pi n_i/L_i\}$, held fixed as the $L_i$ grow.
The sector is spanned by the $N_s$ orthonormal states
$|\bm k\rangle=c^\dagger_{\bm k\uparrow}c^\dagger_{\bm K-\bm k\downarrow}|0\rangle$,
and on it
\begin{equation}
H_L=H_0-ig\,|\beta\rangle\langle\beta|,\qquad
H_0|\bm k\rangle=\varepsilon_2(\bm k)|\bm k\rangle,\qquad
|\beta\rangle=N_s^{-1/2}\sum_{\bm k}|\bm k\rangle ,
\label{eq:sp:HL}
\end{equation}
with $\varepsilon_2(\bm k)=\varepsilon(\bm k)+\varepsilon(\bm K-\bm k)
=-4\sum_i\tau_i\cos(k_i-K_i/2)$ up to the sign of each $\cos(K_i/2)$, and
$\tau_i=t_i|\cos(K_i/2)|$. The loss term is $-ig$ times the projector on the
zero-momentum doublon $|\beta\rangle$, because every matrix element of $D$ in
the sector equals $1/N_s$. Nothing below uses the form of $\varepsilon_2$
except in two places, the quadrature bound~\eqref{eq:sp:quad} and the census
of Eq.~\eqref{eq:sp:census}; everything else uses only that the loss is rank
one.

\paragraph{Two exact identities on every finite lattice.}
Let $H_L\psi=E\psi$. Taking the Hermitian product with $\psi$,
$E\,\langle\psi|\psi\rangle=\langle\psi|H_0|\psi\rangle-ig\,|\langle\beta|\psi\rangle|^2$,
and $\langle\psi|H_0|\psi\rangle$ is real, so
\begin{equation}
\mathrm{Im}\,E=-g\,\frac{|\langle\beta|\psi\rangle|^{2}}{\langle\psi|\psi\rangle},
\qquad
\mathrm{Re}\,E=\frac{\langle\psi|H_0|\psi\rangle}{\langle\psi|\psi\rangle}.
\label{eq:sp:decay}
\end{equation}
Hence $-g\le\mathrm{Im}\,E\le0$ and $|\mathrm{Re}\,E|\le W:=\max_{\bm k}
|\varepsilon_2(\bm k)|=4\sum_i\tau_i$ for every eigenvalue. The decay rate of a
level is $g$ times its weight on the doublon, the two-body form of the identity
$\mathrm{Im}\,E=-g\langle D\rangle$. The second identity is the trace. The sum
of the eigenvalues, with algebraic multiplicity, is
$\mathrm{tr}\,H_L=\sum_{\bm k}\varepsilon_2(\bm k)-ig\,\langle\beta|\beta\rangle$,
so
\begin{equation}
\sum_{n=1}^{N_s}\mathrm{Im}\,E_n=-g
\label{eq:sp:sum}
\end{equation}
exactly, at every $L$ and every $g$: the sector holds one doubly occupied
configuration's worth of loss, and the levels share it. Since no level has
$\mathrm{Im}\,E>0$, the sum rule already bounds how many levels can decay
fast. If $\mathcal N_L(\eta)$ levels have $\mathrm{Im}\,E<-\eta$, their rates
alone exceed $\eta\,\mathcal N_L(\eta)$ and cannot exceed $g$, so
\begin{equation}
\mathcal N_L(\eta)<g/\eta
\label{eq:sp:markov}
\end{equation}
uniformly in $L$. What remains is to show that the bound is one, not $g/\eta$.

\paragraph{The decaying levels are the zeros of the secular function.}
For $z$ off the real axis, $z-H_L=(z-H_0)+ig|\beta\rangle\langle\beta|$ is a
rank-one update of an invertible matrix, and the determinant identity for such
updates gives
\begin{equation}
\det(z-H_L)=\det(z-H_0)\,F_L(z),\qquad
F_L(z)=1+igG_L(z),\qquad
G_L(z)=\frac1{N_s}\sum_{\bm k}\frac{1}{z-\varepsilon_2(\bm k)} .
\label{eq:sp:det}
\end{equation}
Because $\det(z-H_0)$ has no zeros off the real axis, a non-real $z$ is an
eigenvalue of $H_L$ of algebraic multiplicity $m$ exactly when it is a zero of
$F_L$ of order $m$. The real eigenvalues are of a single kind: by
Eq.~\eqref{eq:sp:decay} their eigenvectors are orthogonal to $|\beta\rangle$,
so they are eigenvectors of $H_0$ that never see the loss. In the upper
half-plane $\mathrm{Im}\,G_L(z)=-\mathrm{Im}\,z\,N_s^{-1}\sum_{\bm k}
|z-\varepsilon_2|^{-2}<0$, so $\mathrm{Re}\,F_L=1-g\,\mathrm{Im}\,G_L>1$ and
there are no zeros there, as Eq.~\eqref{eq:sp:decay} requires. Counting
decaying levels is therefore counting zeros of $F_L$ in the lower half-plane.

\paragraph{The finite-lattice secular function converges below the real axis.}
Let
\begin{equation}
G(z)=\int\frac{d^dk}{(2\pi)^d}\,\frac{1}{z-\varepsilon_2(\bm k)},\qquad F=1+igG
\label{eq:sp:G}
\end{equation}
be the infinite-lattice functions of Secs.~\ref{sec:1d}--\ref{sec:hd}. The sum
$G_L$ is the equal-weight rule for this integral on the grid. Split the
Brillouin zone into the $N_s$ cells $\bm k+\prod_i[0,2\pi/L_i)$; in each cell
the integrand differs from its value at the grid point by at most
$\sum_i(2\pi/L_i)\sup|\partial_{k_i}(z-\varepsilon_2)^{-1}|$. With
$|\partial_{k_i}\varepsilon_2|=4\tau_i|\sin(k_i-K_i/2)|\le4\tau_i$ and
$|z-\varepsilon_2|\ge|\mathrm{Im}\,z|$,
\begin{equation}
\sup_{\mathrm{Im}\,z\le-\eta}\big|F_L(z)-F(z)\big|\le
\delta_L(\eta):=\frac{8\pi g}{\eta^{2}}\sum_{i=1}^{d}\frac{\tau_i}{L_i} .
\label{eq:sp:quad}
\end{equation}
The bound is uniform on the whole closed half-plane $\mathrm{Im}\,z\le-\eta$ and
tends to zero as every $L_i$ grows. It is conservative: the integrand is
analytic and periodic, and the equal-weight rule converges geometrically for
such integrands, but the power law is all that the counting needs.

\paragraph{The infinite-lattice secular function has one zero below the axis.}
This is the root census of Sec.~\ref{sec:am}, Eq.~\eqref{thm:am:census}
together with the count on the imaginary axis stated after
Eq.~\eqref{eq:am:gc}, and it is the only place where the hypercubic,
nearest-neighbour band enters. In the sector $\bm K$,
with the effective hoppings $\tau_i$, write $S=\sum_i\tau_ie^{ik_i}$ and
$\Phi(y)=\langle(y^2+16|S|^2)^{-1/2}\rangle$ for $y>0$, the function of
Eq.~\eqref{eq:am:Phi} with $R=|S|$. Then $G(-iy)=i\Phi(y)$,
and
\begin{equation}
\begin{gathered}
F\ \text{has no zero with}\ \mathrm{Im}\,z<0\ \text{except}\ z_*=-iy_*,
\quad g\,\Phi(y_*)=1,\\
\text{present if and only if}\ g>g_c=\big[\langle(4|S|)^{-1}\rangle\big]^{-1}.
\end{gathered}
\label{eq:sp:census}
\end{equation}
Off the imaginary axis the census rules zeros out because
$\mathrm{Re}\,G>0$ on the open fourth quadrant and $\mathrm{Re}\,G<0$ on the
third. On the axis $F(-iy)=1-g\Phi(y)$, and
$\Phi'(y)=-\langle y\,(y^2+16|S|^2)^{-3/2}\rangle<0$, so the zero is unique
and simple, with $F'(z_*)=-ig\Phi'(y_*)\neq0$. Since $\Phi(y)\le1/y$,
also $y_*\le g$. Directions with $\tau_i=0$ drop out of $S$ and leave a
lower-dimensional census; if every $\tau_i$ vanishes the band is flat, $F=1+ig/z$,
and $z_*=-ig$ with $g_c=0$, which the statement already contains.

\paragraph{Counting by Rouch\'e's theorem.}
Fix $g\neq g_c$ and $\eta\in(0,g)$ with $\eta\neq y_*$, and let
$Q_\eta=\{|\mathrm{Re}\,z|<W+g+1,\ -g-1<\mathrm{Im}\,z<-\eta\}$. By
Eq.~\eqref{eq:sp:decay} every eigenvalue with $\mathrm{Im}\,E<-\eta$ lies inside
$Q_\eta$. By Eq.~\eqref{eq:sp:census}, $F$ has no zero on the boundary of
$Q_\eta$, since its only candidate $z_*$ has real part zero, lies above the bottom
edge because $y_*\le g$, and is off the top edge because $\eta\neq y_*$. The
number $m_\eta=\min_{\partial Q_\eta}|F|$ is therefore positive. Both $F$ and
$F_L$ are analytic on a neighbourhood of the closed rectangle, and once
\begin{equation}
\delta_L(\eta)<m_\eta,\qquad\text{i.e.}\qquad
\sum_i\frac{\tau_i}{L_i}<\frac{\eta^{2}m_\eta}{8\pi g},
\label{eq:sp:L0}
\end{equation}
Eq.~\eqref{eq:sp:quad} gives $|F_L-F|<|F|$ everywhere on $\partial Q_\eta$.
Rouch\'e's theorem then says that $F_L$ and $F$ have the same number of zeros
in $Q_\eta$, counted with order. That number is one if $g>g_c$ and $\eta<y_*$,
and zero otherwise. Through Eq.~\eqref{eq:sp:det} it is the number of
eigenvalues with $\mathrm{Im}\,E<-\eta$, which is Eq.~\eqref{eq:sp:main}.
Condition~\eqref{eq:sp:L0} is the explicit $L_0$.

\paragraph{The isolated eigenvalue.}
Apply the same argument to a small disc around $z_*$. On
$|z-z_*|=r\le y_*/2$ Taylor's theorem gives $|F(z)|\ge|F'(z_*)|\,r-\tfrac12M_2r^2$,
with $M_2$ the maximum of $|F''|$ on the disc of radius $y_*/2$, so
$|F|\ge\tfrac12|F'(z_*)|\,r$ for $r\le r_0=\min(y_*/2,|F'(z_*)|/M_2)$. The disc
lies in $\mathrm{Im}\,z\le-y_*/2$, where $|F_L-F|\le\delta_L(y_*/2)$. Choosing
$r_L=4\delta_L(y_*/2)/|F'(z_*)|$, which is admissible once $r_L\le r_0$,
Rouch\'e places exactly one zero of $F_L$ in the disc, and it is the level
counted in Eq.~\eqref{eq:sp:main}. Hence
\begin{equation}
|E_b+iy_*|<\frac{128\pi}{y_*^{2}\,|\Phi'(y_*)|}\sum_{i=1}^{d}\frac{\tau_i}{L_i} .
\label{eq:sp:rate}
\end{equation}
When every $L_i$ is even, $\bm\pi$ lies on the grid and $\Sigma$ of
Eq.~\eqref{eq:sm:sigma} acts within the sector: the shift $\bm k\to\bm k+\bm\pi$ sends $\varepsilon_2\to-\varepsilon_2$
and $|\beta\rangle\to|\beta\rangle$, so $\Sigma H_L\Sigma^{-1}=-H_L$ and the
spectrum is invariant under $E\to-E^*$ with multiplicities. The level $E_b$ is
the only one with $\mathrm{Im}\,E<-\eta$, and $-E_b^*$ has the same imaginary
part, so $-E_b^*=E_b$ and $\mathrm{Re}\,E_b=0$ exactly, on every such lattice.
For $\eta<y_*$ fixed and $L\ge L_0$ it is also the level of most negative
$\mathrm{Im}\,E$ in the sector, which is the property the Letter uses when it
identifies the most-decaying level above threshold with the bound pair.

\paragraph{Where the loss goes.}
The sum rule~\eqref{eq:sp:sum} turns the count into a statement about how
the loss is shared. Above threshold the bound pair carries $y_*$ of the total
$g$ in the limit, and the other $N_s-1$ levels share the remaining $g-y_*$,
an average of $-(g-y_*)/(N_s-1)$ each. Below threshold no level keeps a finite
rate and the $N_s$ levels share all of $g$, an average of $-g/N_s$. The
familiar picture of extended states whose weight at zero separation is of order
$1/N_s$ is thus exact on average. It is not proved level by level.

\paragraph{What the argument does and does not establish.}
Only the census~\eqref{eq:sp:census} uses the lattice. Everything else, the
identities~\eqref{eq:sp:decay} and~\eqref{eq:sp:sum}, the factorisation~\eqref{eq:sp:det},
the convergence~\eqref{eq:sp:quad} and the Rouch\'e count, holds for any lattice
with on-site loss and any band. On any such lattice the number of levels with
$\mathrm{Im}\,E<-\eta$ converges to the number of zeros of $F$ below
$-\eta$, and only that number would have to be recomputed.
Three limitations remain. First, there is no level-by-level rate for the
decaying continuum: Eq.~\eqref{eq:sp:main} says each such level eventually lies
above any fixed $-\eta$, and Eq.~\eqref{eq:sp:sum} says their mean is of order
$1/N_s$, but levels near energies where the finite lattice has large
degeneracies, such as $E=0$ when $4\,|\,L_i$, can approach the axis slowly.
Second, $L_0$ is not uniform in $g$: as $g\to g_c$ both $y_*$ and $m_\eta$ tend
to zero, and at $g=g_c$ there is no gap to protect. Third, the rank-one
structure is special to one pair. With several pairs the trace identity
survives, $\sum_n\mathrm{Im}\,E_n=-g\,\mathrm{tr}\,D$ over the sector, but the
factorisation~\eqref{eq:sp:det} does not, and the most-bound many-pair level
is selected numerically, not by this argument.
The real roots $\pm E_0(g)$ below threshold are consistent with all of this:
they are zeros of the boundary value $F(E-i0)$ on the band, not zeros below the
axis, and by Eq.~\eqref{eq:sp:decay} no eigenvalue can sit on the real axis
while carrying weight on the doublon.
These statements hold on a ring, from exact diagonalization alone
[Fig.~\ref{fig:pt}(d)--(f)]. Every level
of all $L^2$ two-particle states, over every total momentum, obeys
Eq.~\eqref{eq:sp:decay}: the largest deviation, over every level of rings up to
$L=\SPLmax$, is $\SPDone$, so no level enters the upper half-plane. In each
momentum sector the bound pair is identified without the secular equation, as
the most-decaying level whose doublon weight
$|\langle\beta|\psi\rangle|^2/\langle\psi|\psi\rangle$ does not change when $L$ is
doubled. Sectors bind one after another as $g$ passes their own threshold, and at
most one level per sector does so. The second member of the pair that coalesced
at the finite-lattice EP is the slower partner: its doublon weight falls as
$1/L$, its wavefunction extends over the ring, and its decay rate tends to zero,
as Eq.~\eqref{eq:sp:main} requires. In one dimension the analytic continuation of the
lower-lip secular function through the band has a second-sheet root at
$+iy_*$, whereas the physical-sheet equation gives $F(+iy_*)=1+g\Phi(y_*)>0$. In
the relative coordinate $\psi(r)\propto e^{iq|r|}$, and the doublon site fixes
$\sin q=-g/4t$ at $K=0$. Above threshold the two continued solutions have
$q=-\pi/2\pm i\kappa$ with $\cosh\kappa=g/4t$: the bound pair decays as
$e^{-\kappa|r|}$ and has $E=-iy_*$, and the ED level lies on it, while the
$+iy_*$ solution grows as $e^{+\kappa|r|}$. It is a non-normalizable virtual
state, not an eigenvalue of the lossy Hamiltonian.
\endgroup

\section{Bethe ansatz in one dimension}
\label{sec:1d:ba}

Nothing in the resolvent derivation of Sec.~\ref{sec:1d} used integrability. The Bethe ansatz does, and it reproduces every statement of that section independently: the sub-threshold real solutions, the threshold, the bound state, the exponent and the $S$-matrix. The Bethe-ansatz solvability of the Hubbard chain with two-body loss, at the level of the full Liouvillian, was established by Nakagawa \emph{et al.}~\cite{S_Nakagawa2021}; here only the two-body sector of the effective Hamiltonian is needed. For two particles the ansatz is exact and elementary, and the non-Hermitian coupling enters through a single complex rapidity equation. The table at the end of this section summarises the agreement.

\paragraph{Reduction to a relative-coordinate chain.}
Write $\psi(x_1,x_2)=e^{iK(x_1+x_2)/2}\chi(x_1-x_2)$. Each of the four hoppings
carries a centre-of-mass phase $e^{\pm iK/2}$, and the two ways of increasing the
separation by one combine as $e^{iK/2}+e^{-iK/2}=2\cos(K/2)$, so the relative
coordinate obeys a single-impurity chain problem
\begin{equation}
E\,\chi(r)=-J\big[\chi(r+1)+\chi(r-1)\big]+U\,\delta_{r,0}\,\chi(0),
\qquad J\equiv2t|\cos(K/2)|=\tfrac a2 .
\label{eq:1d:rel}
\end{equation}
The triplet ($\chi$ odd) has $\chi(0)=0$ and is untouched by $U$; the singlet
($\chi$ even) is the interacting channel.

\paragraph{Scattering states and the Bethe phase shift.}
Take the even Bethe form $\chi(r)=\cos(q|r|+\delta)$, which solves
Eq.~\eqref{eq:1d:rel} for $r\neq0$ with
\begin{equation}
E=-2J\cos q=-a\cos q .
\end{equation}
The equation at $r=0$ fixes $\delta$. Substituting and expanding
$\cos(q+\delta)=\cos q\cos\delta-\sin q\sin\delta$,
\begin{equation}
-2J\cos q\,\cos\delta=-2J\cos(q+\delta)+U\cos\delta
\;\Longrightarrow\;
0=2J\sin q\,\sin\delta+U\cos\delta ,
\end{equation}
so the two-body phase shift is
\begin{equation}
\boxed{\;\tan\delta(q)=-\frac{U}{2J\sin q}=-\frac{U}{a\sin q}\;}
\label{eq:1d:phase}
\end{equation}
For real $U$ this is the familiar lattice contact-interaction phase shift. For
$U=-ig$ it is complex, which is the whole non-Hermitian content. The $S$-matrix
of the channel is $S=e^{2i\delta}=(1+i\tan\delta)/(1-i\tan\delta)$, i.e.
\begin{equation}
S(q)=\frac{a\sin q-g}{a\sin q+g}=\frac{v-g}{v+g},\qquad v=a\sin q=\sqrt{a^2-E^2},
\label{eq:1d:ba:S}
\end{equation}
which is Eq.~\eqref{eq:1d:smatrix} of the resolvent route, obtained here without ever writing
a Green function. Its zeros, $v=g$, are again $E^2=a^2-g^2$.

\paragraph{Bound states: the rapidity becomes complex.}
Take the decaying Bethe form $\chi(r)=e^{-\kappa|r|}$. For $r\neq0$,
$E=-J(e^{\kappa}+e^{-\kappa})=-2J\cosh\kappa$; at $r=0$,
$E=-2J e^{-\kappa}+U$. Equating the two,
\begin{equation}
-2J\sinh\kappa=U
\qquad\Longrightarrow\qquad
\boxed{\;\sinh\kappa=\frac{ig}{a}\;}
\label{eq:1d:string}
\end{equation}
using $U=-ig$ and $a=2J$. This single equation contains the whole transition.
\begin{itemize}
\item \emph{$g<a$.} Then $\kappa$ is purely imaginary, and there are two
solutions, $\kappa=i\theta$ and $\kappa=i(\pi-\theta)$ with $\sin\theta=g/a$,
$0<\theta<\pi/2$: the envelope does not decay, and neither is a bound state.
Their energies are $E=-2J\cosh\kappa=-a\cos\theta=-\sqrt{a^2-g^2}$ and
$E=-a\cos(\pi-\theta)=+\sqrt{a^2-g^2}$, i.e.\ precisely the two real
sub-threshold roots, which the resolvent route identified as absorption zeros.
The Bethe ansatz thus reproduces both roots as \emph{non-normalisable}
solutions, exactly as it should.
\item \emph{$g>a$.} Write $\kappa=w+i\pi/2$; then
$\sinh\kappa=i\cosh w$, so Eq.~\eqref{eq:1d:string} gives $\cosh w=g/a$, i.e.
$w=\mathrm{arccosh}(g/a)>0$: a genuine decaying envelope
$e^{-w|r|}e^{-i\pi|r|/2}$, whose phase from the $i\pi/2$ has period four in the
relative coordinate $r$. Its energy is
\begin{equation}
E=-a\cosh\!\big(w+\tfrac{i\pi}2\big)=-i\,a\sinh w=-i\sqrt{g^2-a^2} ,
\end{equation}
the bound, decaying pair of the resolvent route.
\item \emph{$g=a$.} Then $\kappa=i\pi/2$ exactly: the two branches meet. The
envelope is oscillatory, so this collision is an exceptional point of the
continued problem, not a normalizable eigenstate of the infinite chain.
\end{itemize}

\paragraph{The exceptional point as a collision of rapidities.}
In terms of the relative rapidity $q$ (with $E=-a\cos q$, so $\kappa=iq$),
the two sub-threshold solutions sit at $q=\theta$ and $q=\pi-\theta$, the two
solutions of $\sin q=g/a$ in $(0,\pi)$, with energies $\mp\sqrt{a^2-g^2}$.
As $g\to a^-$ they move together and collide at the zone value
\begin{equation}
q=\pi/2 \qquad (E=0),
\end{equation}
and for $g>a$ they leave the real axis as $q=\pi/2\mp iw$. The branch
$q=\pi/2-iw$ is the normalisable string $\kappa=w+i\pi/2$ found above, with
$E=-ia\sinh w$; its partner $q=\pi/2+iw$ has a growing envelope and is not an
eigenstate. Near the collision the exact relation $\cosh w=g/a$ gives, with
$\cosh w\simeq1+w^2/2$, the behaviour $w\simeq\sqrt{2(g-a)/a}$, so
\begin{equation}
E\simeq-i\sqrt{2a}\,\sqrt{g-g_c},\qquad\text{and below threshold}\quad
E\simeq\pm\sqrt{2a}\sqrt{g_c-g},
\end{equation}
the same square-root law as Eq.~\eqref{eq:1d:sqrt}. In Bethe language the
exceptional point is therefore a \textbf{square-root branch point in the rapidity
plane}: two Bethe roots collide and acquire an imaginary part, i.e.\ a two-string
forms. The threshold is where the string first becomes normalisable.

\paragraph{Real $U$ has no threshold.}
Equation~\eqref{eq:1d:string} with real attractive $U=-|U|$ reads
$\sinh\kappa=|U|/a>0$, which has a real positive solution for \emph{every}
$|U|>0$: in one dimension a real contact attraction binds at arbitrarily weak
coupling, with $E=-\sqrt{a^2+U^2}$ below the band, and no transition occurs. The
imaginary coupling is qualitatively different: the string acquires the extra
$i\pi/2$, the bound state sits on the imaginary axis at the \emph{centre} of the
band rather than below it, and it exists only for $g>a$. The exceptional point is
the boundary between the two behaviours, and it has no Hermitian counterpart.

\paragraph{Finite ring.}
On a ring of $L$ sites the two-particle Lieb--Wu equations read, for the sign
convention $H=-tT+UD$ of Eq.~\eqref{eq:1d:H}, $T$ the hopping operator, with $u=U/4t$ and one spin
rapidity $\Lambda$,
\begin{equation}
e^{ik_jL}=\frac{\Lambda-\sin k_j-iu}{\Lambda-\sin k_j+iu}\ (j=1,2),
\qquad
\prod_{j=1,2}\frac{\Lambda-\sin k_j-iu}{\Lambda-\sin k_j+iu}=1 ,
\label{eq:1d:liebwu}
\end{equation}
\cite[Ch.~3]{S_Essler2005}. We quote these without re-deriving them. Their
reduction to Eq.~\eqref{eq:1d:string} is immediate: write $k_{1,2}=K/2\pm i\kappa$
with $\mathrm{Re}\,\kappa>0$, so that $e^{ik_1L}\to0$ and $e^{ik_2L}\to\infty$
as $L\to\infty$; the first equation then forces
$\Lambda=\sin k_1+iu=\sin k_2-iu$ (the second equation is the product of the two
first ones, $e^{iKL}=1$), whence $\sin k_1-\sin k_2=2\cos(K/2)\sin(i\kappa)=2i\cos(K/2)\sinh\kappa=-2iu$,
i.e.\ $\sinh\kappa=-U/[4t\cos(K/2)]=-U/a$, which is Eq.~\eqref{eq:1d:string}.
The sign of $iu$ matters: with it reversed the same reduction gives
$\sinh\kappa=+U/a$, for which a real attractive $U$ would have no bound state.
Both the sign and the reduction were checked against exact two-body
diagonalisation on a finite ring at real attractive $U$ and $K=0$: the
bound-state energy of the ring reproduces $e^{ik_1L}$ as given by
Eq.~\eqref{eq:1d:liebwu} to all digits, and not its reciprocal.
Finite-$L$ solutions of Eq.~\eqref{eq:1d:liebwu} with complex $u$ require care
(the string hypothesis is formulated for real $U$), which is a further reason the
paper's arguments are built on the resolvent route.

\paragraph{The two routes compared.}
Table~\ref{tab:routes} sets the resolvent route of Sec.~\ref{sec:1d} beside the
Bethe ansatz, statement by statement.
\begin{table}[h]
\caption{The one-dimensional pair at $U=-ig$ by the resolvent route
(Sec.~\ref{sec:1d}) and by the Bethe ansatz, with $a=4t|\cos(K/2)|$ the
half-width of the relative band and $v=\sqrt{a^2-E^2}$. Both give the same
sub-threshold CPA zeros, the same $S$-matrix, the same threshold and the same
square-root exponent; only the resolvent route generalises to higher
dimensions.}
\label{tab:routes}
\small
\begin{tabular}{p{4.3cm}p{4.9cm}p{4.9cm}}
\toprule
statement & resolvent route & Bethe ansatz \\
\midrule
sub-threshold solutions & real roots of $1=UG(E-i0)$, $E^2=a^2-g^2$ & real rapidities $q=\theta$ and $q=\pi-\theta$, $\sin\theta=g/a$, i.e.\ $E=\mp\sqrt{a^2-g^2}$ \\
what they are & zeros of $s=F(E-i0)/F(E+i0)$ & zeros of $S=e^{2i\delta}$, Eq.~\eqref{eq:1d:ba:S} \\
$S$-matrix & $(v-g)/(v+g)$ & $(v-g)/(v+g)$, identical \\
threshold & $g_c=a=1/\pi\rho_a(0)$ & $g_c=a$: where $\sin\theta=1$, the rapidities collide \\
above threshold & $E=-i\sqrt{g^2-a^2}$, physical sheet & string $\kappa=\mathrm{arccosh}(g/a)+i\pi/2$, same $E$ \\
nature of the transition & double root of the continued $F$ & collision of two Bethe rapidities; branch point in $q$ \\
exponent & $E\simeq\pm\sqrt{2a}\sqrt{g_c-g}$ & identical \\
generalises to $d\ge2$? & yes, this is the route used & no \\
\bottomrule
\end{tabular}
\end{table}

\noindent Numerically, at $t=1$, $K=0$ ($a=4$): $g_c=4$; at $g=3$ both routes give
$E=\pm\sqrt{16-9}=\pm\sqrt7$; at $g=5$ both give $E=-i\sqrt{25-16}=-3i$, and
the Bethe ansatz additionally exhibits it as a string with
$w=\mathrm{arccosh}(5/4)=\ln2$, $a\sinh w=4\times\tfrac34=3$.